\documentclass[prd,showpacs,amsmath,showkeys,twocolumn,floatfix]{revtex4} 

\usepackage{bm}
\usepackage{graphicx}

\usepackage{subfigure}

\usepackage[usenames]{color}
\usepackage[normalem]{ulem}

\def\lsim{\mathrel{\rlap{\lower4pt\hbox{\hskip1pt$\sim$}}
    \raise1pt\hbox{$<$}}}
\def\gsim{\mathrel{\rlap{\lower4pt\hbox{\hskip1pt$\sim$}}
    \raise1pt\hbox{$>$}}}

\def\A{{\bf A}}
\def\D{{\bf D}}
\def\x{{\bf x}}
\def\y{{\bf y}}
\def\k{{\bf k}}
\def\q{{\bf q}}
\def\hx{{\hat x}}
\def\hy{{\hat y}}
\def\hA{{\hat A}}

\begin{document}

\title{A Numerical Approach to Coulomb Gauge QCD}
\author{Hrayr H.~Matevosyan}
\author{Adam P.~Szczepaniak}
\affiliation{Department of Physics and Nuclear Theory Center \\
 Indiana University  Bloomington IN, 47405 USA}
\author{Patrick O.~Bowman}
\affiliation{Centre of Theoretical Chemistry and Physics, Institute of Fundamental
Sciences, \\  Massey University (Auckland), Private Bag 102904, NSMSC, Auckland
NZ}

\begin{abstract}
We calculate the ghost two-point function in Coulomb gauge QCD with a simple
model vacuum gluon wavefunction using Monte Carlo integration. This 
approach extends the previous analytic studies of the ghost propagator with this
ansatz, where a ladder-rainbow expansion was unavoidable for calculating the
path integral over gluon field configurations. The new  approach allows us to
study the possible critical behavior of the coupling constant, as well as
the Coulomb potential derived from the ghost dressing function. We demonstrate
that IR enhancement of the ghost correlator or Coulomb form factor fails to
quantitatively reproduce confinement using Gaussian vacuum wavefunctional.
\end{abstract}

\maketitle

\section{Introduction}

A combination of analytical calculations, based on Dyson-Schwinger 
equations~\cite{{Fischer:2006ub,von_Smekal:1997vx,Braun:2007bx,Zwanziger:2001kw,Pawlowski:2003hq,Lerche:2002ep,Aguilar:2004sw,Frasca:2007uz,Boucaud:2007va,Boucaud:2007hy,Chernodub:2007rn,Dudal:2007cw},Natale:2006nv} and 
 lattice gauge simulations~\cite{Cucchieri:1997dx,Ilgenfritz:2006he,Leinweber:1998im,Furui:2003jr,Cucchieri:2006xi,Cucchieri:1999sz,Cucchieri:2003di,Sternbeck:2005tk,Boucaud:2005gg,Bogolubsky:2005wf,Cucchieri:2006tf,Oliveira:2006zg,Oliveira:2006yw,Fischer:2002eq,Fischer:2005ui,Fischer:2007pf,Fischer:2007mc,Cucchieri:2007md,Bogolubsky:2007ud,Sternbeck:2007ug,Cucchieri:2007zm,Cucchieri:2007rg,Bowman:2007du}, has  
given new insights into the behavior of QCD Green's functions. In 
particular, it has been found that in the Landau gauge at 
low momentum the ghost propagator is enhanced while the gluon propagator 
is suppressed. Dyson-Schwinger equations potentially admit solutions 
that are critical in the infrared (IR), {\it i.e.} the ghost propagator is 
divergent and the gluon propagator vanishes at zero momentum. On the other 
hand, the interpretation  of lattice results is still somewhat controversial, 
since the IR region is sensitive to finite volume effects and possible lattice 
artifacts in mapping between the continuum and lattice definition of propagators~\cite{Fischer:2006ub,Fischer:2002eq,Fischer:2005ui,Fischer:2007pf,Fischer:2007mc}.  One of the original motivations for such studies  follows from the observation that for the physical spectrum to consist  only of color singlet states   it is necessary that the ghost and gluon propagators are critical (in the sense defined above)~\cite{Kugo:1979gm,Kugo:1995km}.  The absence of colored states in the physical spectrum is often taken as a manifestation of confinement.  The relation between 
the IR behavior of the ghost and gluon propagators and the expectation value of the color charge is tied to the realization of the residual gauge symmetry 
remaining after imposing the Landau gauge condition~\cite{Caudy:2007sf}. The 
connection between remnant gauge symmetries and confinement, however, remains 
an unsettled  issue, therefore so does the relation between the IR behavior of 
the propagators and confinement.  The relationship between the gluon and 
ghost propagators and confinement can be investigated in other gauges, and the 
Coulomb gauge can be particularly illuminating~\cite{Zwanziger:2002sh,Zwanziger:1995cv,Cucchieri:1996ja,Cucchieri:2006hi,Epple:2006hv,Epple:2007ut}.  

In the Coulomb gauge the time component of the vector potential becomes 
constrained by the transverse gluon field defined in the spatial directions 
alone.   $\A^a(\x)$ satisfies, $\bm{\nabla} \cdot\A^a = 0$ for all color 
components, $a=1\cdots N_C^2-1$, leading to an instantaneous potential between 
color charges. This potential depends on the inverse of the Faddeev-Popov, or 
ghost, operator, $M^{-1}(A) = [\bm{\nabla} \cdot \D(A)]^{-1}$, with $\D(A)$  
being the covariant derivative in the adjoint representation. It was postulated
by Gribov~\cite{Gribov:1977wm} and by Zwanziger~\cite{Zwanziger:1993dh} that 
gauge field configurations near the boundary of the field space domain, the 
Gribov horizon, dominate matrix elements and, since at the boundary the 
Faddeev-Popov operator vanishes, the instantaneous Coulomb potential is 
expected to be enhanced compared to the value at zero field~\cite{Gribov:1977wm,Zwanziger:1993dh,Dokshitzer:2004ie,Zwanziger:1990by,Zwanziger:1991gz}.  
This could signal confinement.  Furthermore, since for a state containing a 
static quark-antiquark pair in the vacuum the Coulomb potential provides an 
upper limit on the total energy, Zwanziger concluded that a necessary condition
for confinement is that the expectation value of the Coulomb potential in such 
state is also confining~\cite{Zwanziger:2002sh}.  
From the point of  view  that the energy spectrum is a direct probe of
confinement, it seems relevant to investigate matrix elements of  the inverse 
of the Faddeev-Popov operator. Analytical calculations have been performed in, 
for example, Refs.~\cite{Cucchieri:1996ja,Epple:2006hv,Epple:2007ut,Szczepaniak:2001rg,Szczepaniak:2003ve,Feuchter:2004mk,Feuchter:2004gb,Reinhardt:2004mm}. These typically start from an ansatz for the vacuum wave functional and various approximations are used to derive Dyson equations for correlations functions. Since the Coulomb energy involves fields at one time slice, only spatial correlations are needed. 
Through a  systematic study of the IR behavior of  the  gluon-gluon correlation function and the Faddeev-Popov operator it  was shown that  within the particular set of  approximations used to derive the Dyson equations, all 
self consistent solutions are IR finite, but close to being critical. Most 
likely what this means is that the vacuum wave functionals used in these calculations do not yet account for all field configurations responsible for confinement. Another way of seeing this is through the behavior of the spatial Wilson 
loops, for which such  wave functionals  fail to reproduce the area law behavior.  If and when missing configurations are properly accounted for one would still face the question of reliability regarding the other approximations used in 
deriving the Dyson equations. These are typically based on the large-$N_C$ expansion and examination of the IR and ultraviolet (UV)  behavior of higher order 
diagrams.  To leading order this amounts to summing the rainbow-ladder diagrams. 
       
       In this paper we confront the Dyson equations for the Coulomb  gauge   correlators  with direct evaluation of the underlying  matrix elements  using  Monte Carlo  techniques  for the path integral over the transverse gluon fields. The numerical techniques are  close in spirit to those  of lattice gauge theory, 
and are detailed in Section~\ref{MCCalculation}.  We begin by giving, in 
Section~\ref{CoulombGauge}, a short summary of the Coulomb gauge and 
derivation of the Dyson equation.  A summary and conclusions are given in 
Section~\ref{Summary}.

\section{Coulomb Gauge QCD}
\label{CoulombGauge} 

In the Schr\"{o}dinger representation the degrees of freedom of the Coulomb 
gauge Yang-Mills theory are: the transverse gluon fields, $\A^a(\x)$, which are
the generalized coordinates, and their conjugate momenta
$\bm{\Pi}^a(\x) = -i \delta/\delta \A^a(\x)$, equal to the negative of the
 transverse chromo-electric field~\cite{Christ:1980ku}.  These satisfy
the canonical commutation relation, 
\begin{equation}
\lbrack \Pi ^{i,a}(\mathbf{x}),A^{j,b}(\mathbf{y})]=-i\delta _{ab}\delta
_{T}^{ij}(\mathbf{\nabla }_{\mathbf{x}})\delta (\mathbf{x}-\mathbf{y}),
\end{equation}%
where $\delta _{T}^{ij}$ is the transverse projector $\delta _{T}^{ij}(%
\mathbf{\nabla })=\delta _{ij}-\nabla _{i}\nabla _{i}/\mathbf{\nabla }^{2}$.
The canonical Hamiltonian is a function of the generalized coordinates and 
momenta, and is given by 
\begin{equation}
H=\frac{1}{2}\int d\mathbf{x}\left[ {\mathcal{J}}^{-1}\mathbf{\Pi }^{a}(%
\mathbf{x}){\mathcal{J}}\cdot \mathbf{\Pi }^{a}(\mathbf{x})+\mathbf{B}^{a2}(%
\mathbf{x}) \right] +V \label{H} ,
\end{equation}%
where the  chromo-magnetic field, ${\bf B}$, is given by, 
\begin{equation}
\mathbf{B}^{a}(\mathbf{x})=\mathbf{\nabla }\times \mathbf{A}^{a}(\mathbf{x})+%
\frac{g}{2}f^{abc}\mathbf{A}^{b}(\mathbf{x})\times \mathbf{A}^{c}(\mathbf{x}%
).
\end{equation}%
As usual, repeated indices are summed over.
In Eq.~(\ref{H}), ${\mathcal{J}}=\det (M(A))$ represents the curvature of the 
Coulomb gauge field domain and is given by the Jacobian of the transformation 
from the $A^0=0$ (Weyl) gauge -- which has a flat field space -- to the
Coulomb gauge.  Here, $M$ is the Faddeev-Popov operator, 
\begin{equation}
M^{ab}(\x,\y)=\left[ -\nabla ^{2}_\x\delta ^{ab}+gf^{abc}\mathbf{A}^{c}\cdot \mathbf{%
\nabla}_\x\right]\delta^3(\x-\y).
\end{equation}
The Coulomb potential, $V$, is obtained by using the equations of motion to 
eliminate the longitudinal gauge field, and can be written 
\begin{equation}
V=\frac{1}{2}\int d^{3}\mathbf{x}d^{3}\mathbf{y}{\mathcal{J}}^{-1}\rho ^{a}(%
\mathbf{x}){\mathcal{J}}K^{ab}(\mathbf{x},\mathbf{y};\mathbf{A})\rho ^{b}(%
\mathbf{y}), 
\end{equation}%
where, in the absence of quarks,  the color charge density is given by 
\begin{equation}
\rho ^{a}(\mathbf{x})=f^{abc}\mathbf{\Pi }^{b}(\mathbf{x})\cdot \mathbf{A}%
^{c}(\mathbf{x}),
\end{equation}%
and the  Coulomb kernel , $K(A)$ is 
\begin{equation}
K(A)=gM^{-1}(A)(-\nabla ^{2})gM^{-1}(A).
\end{equation}%
In the abelian limit this kernel reduces to, 
\begin{equation}
K^{ab}(\mathbf{x},\mathbf{y})=\frac{g^{2}\delta ^{ab}}{4\pi |\mathbf{x}-%
\mathbf{y}|},
\end{equation}
the familiar expression for the Coulomb potential between charges located at 
points $\x$ and $\y$. 
Denoting the vacuum wave functional by $\Psi[A] = \langle A| \Psi\rangle $,  
the vacuum expectation value,  ({\it vev}) of an operator 
$\mathcal{O}[A]$ in the Coulomb gauge is given by, 
\begin{equation}
\langle \mathcal{O}\rangle =\frac{\langle \Psi |\mathcal{O}|\Psi \rangle }{%
\langle \Psi |\Psi \rangle },
\end{equation}%
where%
\begin{equation}
\langle \Psi |\mathcal{O}|\Psi \rangle =\int_{\Lambda}  \mathcal{D}A{\mathcal{J}}[A]\,%
\mathcal{O}[A]|\Psi \lbrack A]|^{2},  \label{EQ_FUNC_INT}
\end{equation}
and the integral is restricted to the fundamental modular region (FMR)  $\Lambda \in \Omega$ which is inside the Gribov region $\Omega$. 
The FMR is defined as the set of gauge fields $\A^a(\x)$ corresponding to the 
absolute minima of the functionals $I[g] \equiv \int d\x (\A^{ag}(\x))^2 $ 
minimized with respect to time-independent gauge transformations $g = g(\x)$, 
while the Gribov region $\Omega$ also includes local minima of $I$.  It has 
been argued by Zwanziger~\cite{Zwanziger:2003cf} that the bulk of the integral 
measure is concentrated on the common boundary of FMR and the Gribov region and
in the Monte Carlo simulations presented here only the restriction to $\Omega$ will be 
implemented. 
  The {\it vev} of the  inverse of the Faddeev-Popov operator, which in the 
Coulomb gauge plays the dual role of the ghost propagator and the running 
coupling, is given by 
\begin{equation}
\frac{d(k)}{k^{2}}=\frac{1}{N_{c}^{2}-1}\delta ^{ab} \int d\x e^{i\k\cdot \x} \langle \Psi | g M^{-1,ab}(\x,{\bf  0}) |\Psi\rangle,   \label{EQ_GHOST_INV_FP}
\end{equation}
where $d(k)$ is referred to as the ghost dressing function; at tree-level, 
$d(k)=1$.  If the expectation value of the Coulomb kernel is approximated by 
the square of the {\it vev} of the ghost propagator then the
momentum space Coulomb potential between a color-singlet static 
quark-antiquark pair becomes $V(k) = -C_F d^2(k)/k^2$~\cite{Zwanziger:1995cv,Epple:2006hv,Cucchieri:1996ja}. In general, however, one expects the two 
{\it vevs} to be different and this difference can be accommodated via an additional form factor and    results  in the potential of the form $V(k) = -C_F d^2(k) f(k)/k^2$~\cite{Swift:1988za,Szczepaniak:2001rg}. It is clear that if the ghost becomes IR enhanced, $d(k) >> 1$ as $k \to 0$, the Coulomb interactions between color charges becomes stronger as the separation between charges increases. To obtain a linearly rising potential, however,  it would be necessary  for the product $d^2(k) f(k)$ to be critical with $d^2(k) f(k) \to k^{-2}$ as $k \to 0$.

\subsection{ Dyson equations} 
The set of coupled  Dyson equations for the ghost dressing function $d(k)$, the  Coulomb dressing function $f(k)$ and the gap equation, which determines the gluon-gluon correlation function, were derived and extensively studied in
Refs.~\cite{Epple:2007ut,Szczepaniak:2001rg,Szczepaniak:2003ve,Feuchter:2004mk,Feuchter:2004gb,Reinhardt:2004mm}. Here we only summarize the main features of the ghost and gluon correlation functions. 
In these studies the vacuum wave functional was parametrized as a gaussian 
\begin{equation} 
\Psi \lbrack A]=\exp\left( {-\frac{1}{2}\int \frac{d^{3}\mathbf{k}}{(2\pi )^{3}}%
\omega (k)\A^{a}(\k) \A^a(-\k)}\right),  \label{EQ_MODEL_WF}
\end{equation}%
with $\omega(k)$ being a parameter. It was shown in 
Refs.~\cite{Szczepaniak:2003ve,Feuchter:2004mk,Feuchter:2004gb,Reinhardt:2004mm} 
that, to leading order in the loop expansion, the effect of the curvature 
${\cal J}$ could be absorbed by a redefinition of $\omega$ with the gap equation correlating the low-mometum behavior of $\omega$ and the curvature.  In the 
subsequent derivations of the Dyson equations we thus set ${\cal J} = 1$  
The vacuum wave functional can be optimized by minimizing the vacuum energy 
density with respect to $\omega(k)$.  This leads to a gap equation which after 
renormalization depends on the renormalized coupling $g_r(\mu)$ and the 
boundary condition $\omega(k\to 0 )= m_g$. As long as $m_g$ is finite one finds
that the solution of gap equation is qualitatively insensitive to $g_r(\mu)$ 
and can be well describe by, 
\begin{equation}
\omega (k)=\biggl\{%
\begin{array}{ll}
m_{g} & \text{if $k<$}m_{g} \\ 
k & \text{otherwise.}%
\end{array}
\label{EQ_OMEGA}
\end{equation}%
\begin{figure}[ptbh]
\begin{center}
\includegraphics[width=0.45\textwidth]{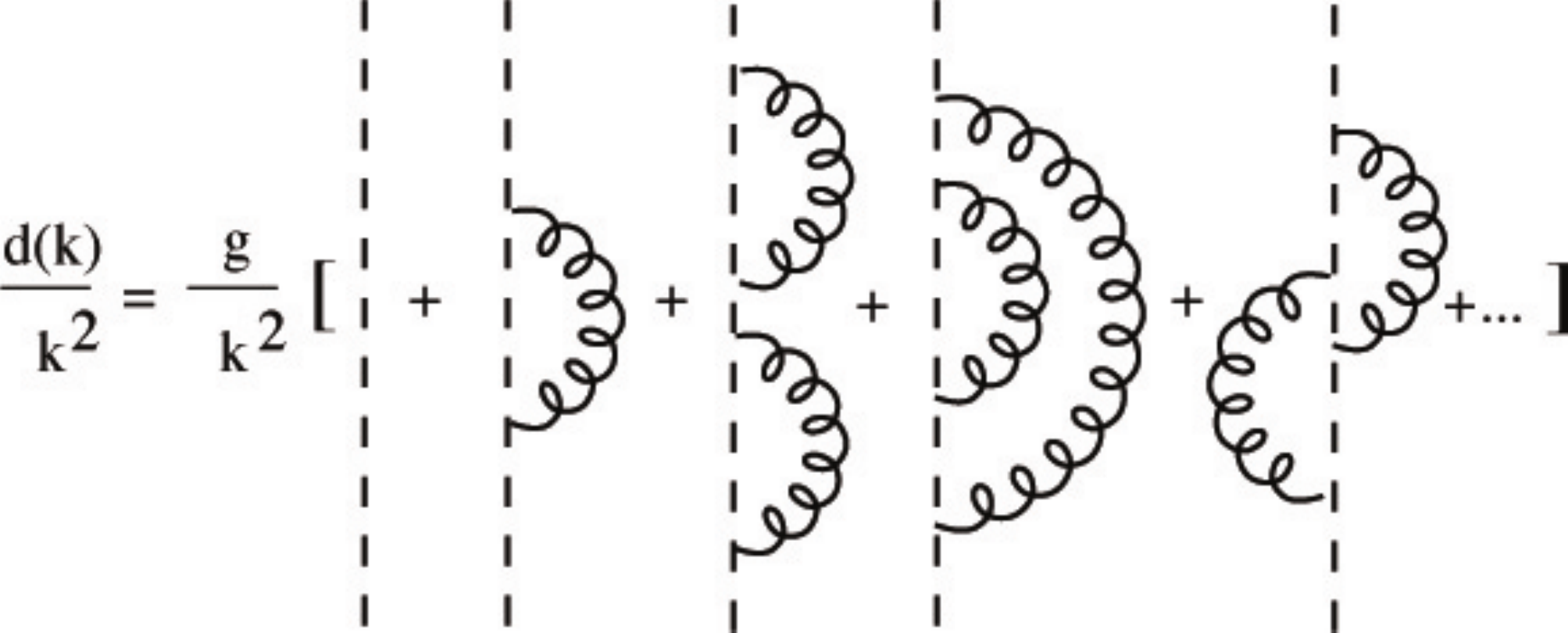}
\end{center}
\caption{Diagrammatic representation of the expansion of the functional integral for the ghost propagator {\it c.f.} Eq.~(\ref{EQ_INV_FP_EXPANSION})} 
\label{dd}
\end{figure}
It should be noted that $m_g$ is a mass parameter introduced by the ansatz wave
functional and should not be confused with the QCD scale introduced by 
renormalization. The latter appears in the renormalized Dyson equation
for the ghost dressing function which, as mentioned earlier, can be identified 
with the running coupling.   In principle, 
$m_g = m_g(g_r(\mu),\mu)$ should be renormalization point invariant and  just like $g_r(\mu)$ determined by a physical observable {\it e.g.} the string tension.  Within the set of truncations build in the derivation of the Dyson series, most likely the   renormalization group invariance of $m_g$ can not be proven and we shall  consider  $m_g$ as a free parameter.  Given $\omega(k)$ the Dyson  series for the ghost dressing function can be  sum up and represented as a single integral equation within the rainbow-ladder approximation, illustrated in Fig.~\ref{dd}. All omitted diagrams have at least one vertex loop correction ({\it e.g.} last diagram in Fig.~\ref{dd}), which were shown to be generally smaller than the self-energy loops~\cite{Szczepaniak:2001rg}. The diagrams shown in Fig.~\ref{dd} represent functional integrals over $|\Psi[A]|^2$ of polynomials of the $A$ field originating from the expansion of the inverse Faddeev-Popov operator
\begin{eqnarray} 
&& {1 \over {\langle \Psi|\Psi \rangle}} \int_\Omega {\cal D} A g {{g |\Psi[A]|^2} \over {-\bm{\nabla}\cdot\D[A]}} \to  \nonumber \\
& \to & {g\over {-\bm{\nabla}^2}} \left[ 1 + {1\over {\langle |\Psi|\Psi\rangle}} \int_\Omega {\cal D} A g \A \times {{\bm{\nabla} } \over {\bm{\nabla}^2}} 
g \A \times {{\bm{\nabla} } \over {\bm{\nabla}^2}}  + \cdots \right] \nonumber \\ \label{EQ_INV_FP_EXPANSION} 
\end{eqnarray}
where $\times$ refers to the color space. Neglecting the restriction to the 
Gribov region enables one to perform the functional integrals analytically, 
and neglecting contractions that corresponds to vertex corrections
makes it possible to re-sum the series, resulting in, 
\begin{equation} 
{1\over {d(k)}}  =  {1\over {g(\Lambda)}}
  - {N_C} \int^\Lambda {{d\q} \over {(2\pi)^3 2\omega(q)}} {{1-(\hat\k\cdot\hat\q)^2} \over {(\k-\q)^2}} d(|\k-\q|) \label{lambda}.
\end{equation}
The dependence of the bare coupling, $g=g(\Lambda)$, and the loop 
integral on the UV cut-off has been shown explicitly.   Instead of using the 
bare coupling and the UV cutoff as the renormalization point, the equation can 
be renormalized at a finite momentum scale through subtraction, which also 
defines the renormalized coupling as $g_r(\mu) \equiv d(k=\mu)$ 
 \begin{eqnarray}
&&{1\over {d(k)}}  =  {1\over {d(\mu)}} \nonumber \\ 
  &- &  {N_C} \int^\Lambda {{d\q} \over {(2\pi)^3}}  \left[ {{1-(\hat\k\cdot\hat\q)^2} \over {(\k-\q)^2}} {{d(|\k-\q|)}\over {2\omega(q)}}  - (|\k| \to \mu) \right].
\nonumber \\ \label{d} 
\end{eqnarray}
\begin{figure}[ptbh]
\begin{center}
\includegraphics[width=0.5\textwidth]{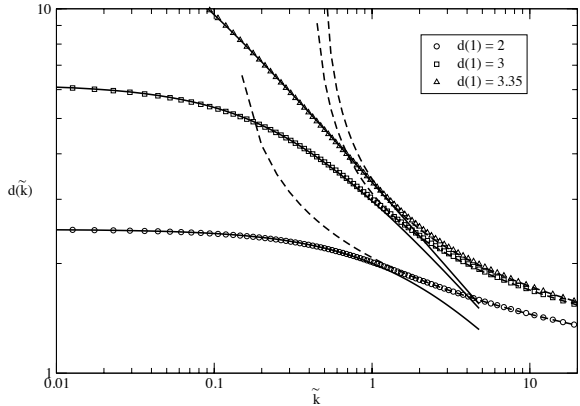}
\end{center}
\caption{Comparison between the numerical solutions of the Dyson equation for 
the ghost propagator and the approximate analytical solutions of 
Eqs.~(\ref{EQ_GHOST_REN_LR_LOW}), (\ref{EQ_GHOST_REN_LR_HIGH}).  The solid 
(dashed) lines represent the low (high) momentum behavior, respectively.} 
\label{d-sd}
\end{figure}
As discussed above, the mass scale is brought in through the function $\omega$,
and in the case discussed here, it is given by $m_g = \omega(0)$.  Thus from 
now on we will use the notation $\tilde k \equiv  k/m_g$ to denote 
dimensionless momenta. The solution of  the Dyson equation for the ghost 
propagator  depends on one more parameter, the value of $d(\tilde \mu)$ at a single point, {\it i.e.} at $\tilde \mu = \mu/m_g = 1$.   In Fig.~\ref{d-sd} we plot the numerical solutions of Eq.~(\ref{d}), as a function of momentum in units 
of $m_g$, for three choices of 
$d(\tilde k=1)$. As $d(1)$ is increased the solutions become more IR enhanced until at,  approximately, $d(1) \sim 3.41$ the solution becomes critical~\cite{Szczepaniak:2001rg}. Above this critical point the Dyson equation has no solutions, {\it i.e.} develops a Landau point at physical, $k>0$ momentum. This is 
a sign that the functional integration in Eq.~(\ref{EQ_INV_FP_EXPANSION}) has crossed the Gribov horizon. The mass scale dependence of the ghost propagator can be best understood by using an angular approximation, to the integral in Eq.~(\ref{d}), 
\begin{equation}
|\k -\q| \to \theta(k - q) k + \theta(q-k) q,
\end{equation} 
which enables one to transform the integral equation to a first order differential equation that can be solved analytically and further well approximated by~\cite{Szczepaniak:2001rg}, 
\begin{eqnarray}
d(\tilde k) &=&\frac{d(\tilde \mu)}{\left[ 1+\beta_L d^{1/\gamma}(\tilde\mu )\left( \tilde k-\tilde \mu \right) \right]^{\gamma}}%
,~\tilde k \leq 1,  \label{EQ_GHOST_REN_LR_LOW} \\
d(\tilde k) &=&\frac{d(\tilde \mu )}{\left[ 1+\beta_Hd^{1/\gamma}(\tilde \mu )\log \left( \frac{\tilde k}{\tilde \mu}\right) \right]^\gamma},
~\tilde k\geq 1,  \label{EQ_GHOST_REN_LR_HIGH}
\end{eqnarray}%
where $\gamma=1/2$ and $\beta_L = (5/6) (N_C/3)/\pi^2$ and $\beta_H =  (N_C/3)/\pi^2 $ {\it i.e.} $\beta_L \sim \beta_H \sim 0.1$ for $N_C=3$. 
It clearly follows that  the ghost propagator is independent of the 
renormalization scale, $\mu$ and depends on a single value of 
$d(\tilde \mu)$ at an arbitrarily chosen renormalization point. Furthermore, 
from Eq.~(\ref{EQ_GHOST_REN_LR_LOW}) it follows that a solution exists, 
{\it i.e} there is no Landau pole, as long as 
$d(\tilde \mu) < 1/(\beta_L \tilde\mu)^\gamma$. 

As discussed above, the approximations leading to Eq.~(\ref{d}) include 
eliminating all vertex corrections and neglecting the restriction on the 
functional integral to be contained within the Gribov horizon.  In the 
following we present results from a Monte Carlo simulation of the ghost 
propagator that does not have these limitations. 


\section{Monte Carlo Calculation}
\label{MCCalculation} 

The evaluation of the functional integral in Eq.~(\ref{EQ_FUNC_INT}) is
usually performed analytically by expanding the operator $\mathcal{O}$ in
a power series over the gauge field $A$ and truncating at some order 
({\it c.f.}  Eq.~(\ref{EQ_INV_FP_EXPANSION}) ). Here we avoid these 
approximations by evaluating the functional integral by Monte Carlo integration
using the model wavefunction~(\ref{EQ_MODEL_WF}) with the approximate solution 
(\ref{EQ_OMEGA}) for $\omega (k)$.

The gluon configurations are generated on a $N_{d}=3$ dimensional momentum
space grid. The gluon fields $A_{i}^{c}(k)$ are $N_{d}\times (N_{C}^{2}-1)$
complex numbers per lattice site, where $A_{i}^{a}(-k)={A_{i}^{a}}^{\ast
}(k) $. The momentum is discretized on the lattice as%
\begin{gather}
k_{i}=\frac{2\pi n_{i}}{aN_{i}}\quad \left\vert n_{i}\right. \quad \in
\left( -\frac{N_{i}}{2},\frac{N_{i}}{2}\right] , \\
i\in \{1,2,3\}, \label{kdis}
\end{gather}%
where $a$ denotes the lattice spacing. The gauge fields must satisfy the position space Coulomb gauge condition 
\begin{equation}
\sum_{i=1}^{3}A^a_{i}(x)-A^a_{i}(x-a\hat{\imath})=0,
\end{equation}%
which translates in the momentum space to%
\begin{equation} 
\sum_{i=1}^{3}(1-\cos (ak)+i\sin (ak)) A^a_{i}(ak)  =  0. 
\label{EQ_GAUGE_INV}
\end{equation} 
From now on we will use the notation $\hat A(\hat k) \equiv A(ak)/a^2$, {\it etc.}  in reference to dimensionless  quantities scaled with the lattice spacing. 
The coupling is incorporated by generating $g\hat A_{i}^{a}(\hat k)$ rather than $%
\hat A_{i}^{a}(\hat k)$, which requires substituting $\omega (k)$ with $\omega
(k)/g^{2}$ in the model wavefunction. The gluon fields are generated with
the distribution 
\begin{equation}
|\Psi \lbrack A]|^{2}=\exp \left\{ -{1\over {N_{L}^3}} \sum_{n_i}\sum_{i=1}^{N_{d}}%
\sum_{a=1}^{N_{c}^{2}-1}\hat A_{i}^{a}(\hat k)\hat A_{i}^{a}(-\hat k)\frac{\hat \omega (\hat k)}{g^{2}}%
\right\} . \label{latwf} 
\end{equation}%
This is accomplished by independently generating two of the vector
components, $\hat A^a_{i}$, with a heatbath, then constructing the third component
such that the momentum space Coulomb gauge condition, Eq.~(\ref{EQ_GAUGE_INV}), is satisfied.

The calculation of the Jacobian is akin to the calculation of the quark
determinant in lattice QCD and in the present work it is set to one. The
Jacobian was included in Ref.~\cite{Szczepaniak:2003ve} in a certain
truncation scheme. There it was found to lessen the dependence of the
Coulomb potential to the choice of coupling.

\begin{figure}[tbph]
\begin{center}
\includegraphics[width=0.4\textwidth
]{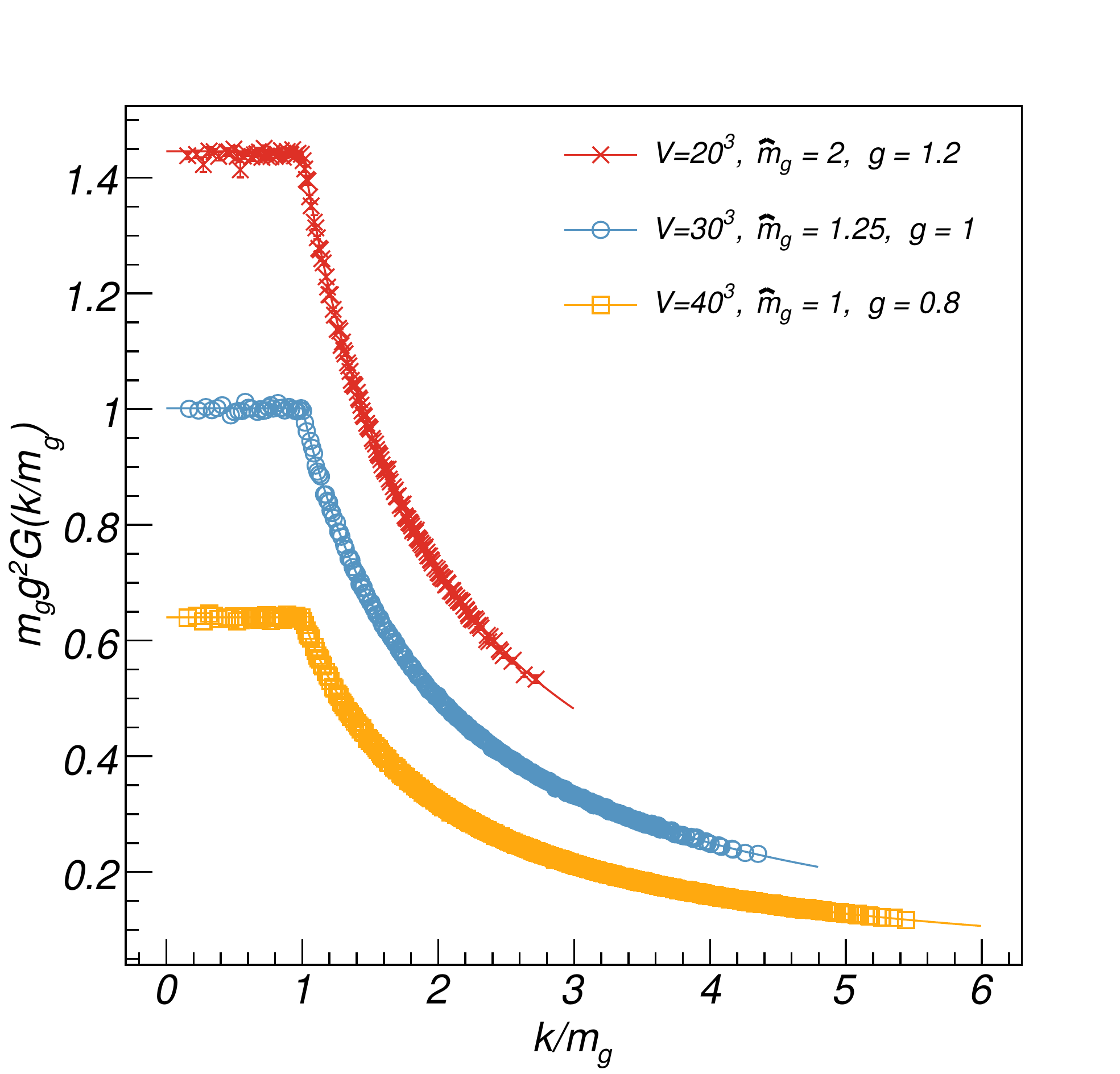}
\end{center}
\caption{(Color Online)
The gluon propagator calculated on $1000$ gauge field configurations with 
various parameters. The analytic results are plotted as a continuous line, 
showing a good agreement with simulations.}
\label{PLOT_GLUON}
\end{figure}

As a first test we evaluate the gluon propagator, 
\begin{equation}
g^{2}\hat G(\hat k)=\frac{1}{N_{L}^3 }\frac{1}{N_{d}-1}\frac{1}{N_{c}^{2}-1}\langle
\sum_{i=1}^{N_d-1} \sum_{a=1}^{N^2_C-1}\hat A_{i}^{a}(\hat k)\hat A_{i}^{a}(-\hat k)\rangle .
\end{equation}%
The value of $G(k)$ is analytically known to be $G(k)=1/2\omega (k)$.
The numerical result, shown in Fig.~\ref{PLOT_GLUON} does indeed agree with
the analytical one, where the numerical statistics are improved by taking
the $Z_{3}$ average, that is, averaging over the three equivalent directions
in momentum space.  Since $\hat k = \hat m_g \tilde k$ 
The physical propagator in units of $m_g$ is given by 
\begin{equation} 
m_g G(k) =  \hat m_g \hat G(\hat k). 
\end{equation} 
We now proceed to computing the ghost dressing
function. The ghost propagator is expressed as the expectation value of the
inverse of the Faddeev-Popov (FP) operator, Eq.~(\ref{EQ_GHOST_INV_FP}). The
discrete form of the FP operator was derived in Ref. \cite{Zwanziger:1993dh} 
\begin{eqnarray}
&& M^{ab}(\hx,\hy){\hat \phi} ^{b}(\hy) =\sum_{i=1}^{N_{d}}\delta ^{ab}\left( {\hat \phi}^{b}(\hy+
\hat{\imath})+{\hat \phi}^{b}(\hy-\hat{\imath})-2{\hat \phi} (\hy)\right)
\label{EQ_FP_DISCRET} \nonumber \\
& &  -f^{abc}\left( {\hat \phi} ^{b}(\hy+\hat{\imath})\hA_{i}^{c}(\hy)-{\hat \phi} ^{b}(\hy-\hat{
\imath})\hA_{i}^{c}(\hy-\hat{\imath})\right),  \notag
\end{eqnarray}
which is real and symmetric. Note that the region of integration in Eq.~(\ref%
{EQ_FUNC_INT}) is the Gribov region where $M$ is positive definite. Thus any 
gauge field configuration that produces a Faddeev-Popov operator with negative 
eigenvalues must be discarded.

With periodic boundary conditions imposed on the lattice, $M^{ab}(x,y)$ has $%
N_{c}^{2}-1$ trivial zero modes, making it formally non-invertible. This
problem is avoided by following Ref.~\cite{Boucaud:2005gg} and solving 
\begin{equation}
\int_{V} d\y M^{ab}(\x,\y){\phi} ^{b}(\y)=\delta ^{ab}\left( \delta (\x)-\frac{1}{V}\right) .
\end{equation}%
The position-color vectors are then Fourier transformed to momentum space
and the inverse of $M^{ab}(x,0)$ recovered. 
\begin{widetext} 
\begin{equation}
\int_Vd\x e^{-i\mathbf{k}\cdot \mathbf{x}}\left\langle \phi
^{a}(x)\right\rangle  =\int_V d\x e^{-i{\mathbf{k}\cdot \mathbf{x}} 
}\left\langle (M^{-1})^{aa}(\x,{\bf 0})\right\rangle -\frac{1}{V}\int_V  d\x d\y e^{-i{ 
\mathbf{k}\cdot \mathbf{x}}}\left\langle
(M^{-1})^{aa}(\x,\y)\right\rangle 
 ={{D(k)}\over g}-\delta ({\bf k})\int_Vd\x {{D(x)} \over g}, 
\end{equation}
\end{widetext}
where $D(k) \equiv d(k)/k^2$. 
The average is taken over gauge field configurations, and finally the ghost
propagator is $Z_{3}$ averaged. In the free case, $M\rightarrow -\nabla ^{2}$
and the propagator would be 
\begin{equation} 
\hat D(\hat k)  =\frac{g}{4\sum_{i}\sin ^{2}( \hat k_i/2)} \equiv \frac{g}{k^2},
  \label{EQ_DEFINE_Q} \\
\end{equation}%
which defines an appropriate momentum variable. The same philosophy is used in
\emph{conventional} lattice QCD studies of gluon propagator 
\cite{Bonnet:2001uh,Leinweber:1998im,Marenzoni:1994ap}.

With the model wavefunction (\ref{EQ_MODEL_WF}), the coupling $g$ is a free
parameter. The larger $g$ is chosen to be, the broader the Gaussian. This
increases the fluctuations of the gauge fields and $M$ develops smaller
eigenvalues, resulting in the infrared enhancement of $\left\langle
M^{-1}(k)\right\rangle $. With increasing the value of $g$, the FP operator
becomes likely to develop negative eigenvalues. While this means that a
(possibly large) proportion of the generated gauge fields must be rejected,
it is necessary for the entire domain of the functional integration to be
sampled. The number of rejected configurations grows rapidly when the value
of $g$ approaches certain critical value, which depends on the value of $%
m_{g}$ used in the model for $\omega (k)$ of Eq.~(\ref{EQ_OMEGA}). 
This is easily understandable, as larger value of $m_{g}$ means the gluon
wavefunction is infrared enhanced in a larger interval of momenta, yielding
narrower Gaussian width over that interval. 

Each generated gauge configuration used in calculating the ghost dressing
function is checked to lay in the Gribov region by calculating several eigenvalues
of the  FP operator to ensure their positivity. If the latter constrain is not imposed,
the resulting ghost propagators are dominated by numerical
fluctuations (resemble random noise) in the region where the generated gauge
configurations have large fraction laying outside of Gribov region. For example,
for $\hat m_{g}=1.5$, the fraction of rejected configuration
ranges from  nearly $0\%$ for $g<1$ to $100\%$ for $g>1.1$ with sharp increase above
$g=1$. For $\hat m_{g}=5$ this ``critical'' value of $g$ increases to about $1.6$ . 
In our calculations we restrict to the region of $g$, where the
fraction of rejected configurations does not exceed $20\%$ to maintain moderate computational time.  
The resulting ghost dressing function is shown in Fig.~\ref{PLOT_GHOST_DIAG} for
a calculation with $1000$ gluon configurations on a $40^{3}$ lattice with $%
\hat m_{g}=1.25$ and $g=0.7$.

 In order to relate the calculated ghost dressing function to the physical
region several issues should be resolved that would allow to draw a
correspondence. Here we review the most relevant ones.

\subsection{Lattice Artifacts}

Discretization of space introduces several artifacts, that should be
accounted for. These are errors introduced by finite lattice volume, finite
lattice spacing, which also induces broken spatial rotational symmetry.

\subsubsection{Finite Lattice Spacing}

\begin{figure}[ptbh]
\begin{center}
\includegraphics[width=0.4\textwidth
]{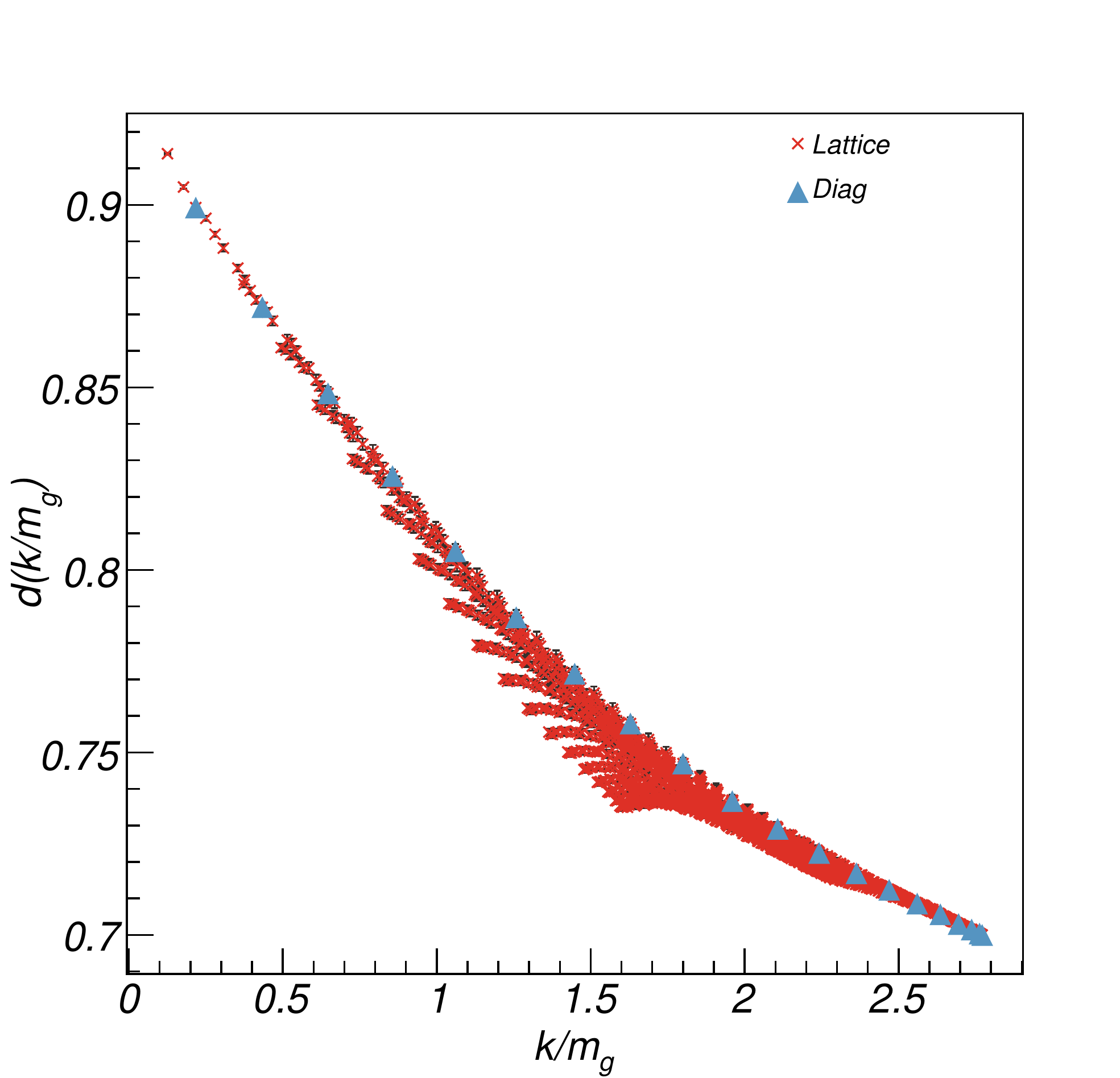}
\end{center}
\caption{(Color Online)
Ghost dressing function $d(k/m_{g})$ versus $k/m_{g}$ for $%
\hat m_{g}=1.25$ and $g=0.7$ on a lattice with $40^{3}$ volume. Here the crosses
denote the full data set and open triangles denote the subset of points with
equal momenta components.}
\label{PLOT_GHOST_DIAG}
\end{figure}

\begin{figure}[ptb]
\centering 
\subfigure[] {
\includegraphics[width=0.4\textwidth]{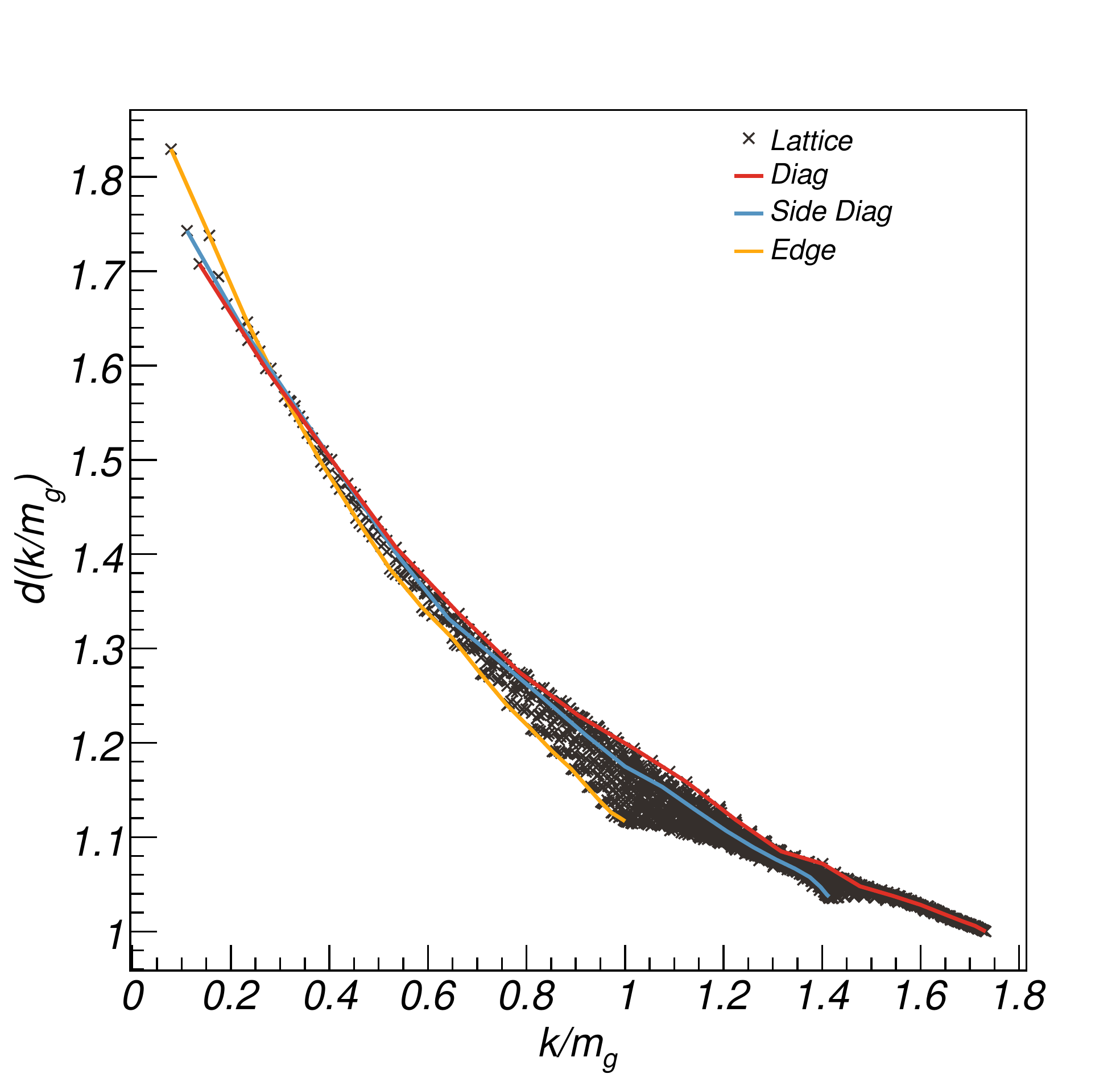} }
\hspace{0.1cm} 
\subfigure[] {
\includegraphics[width=0.4\textwidth]{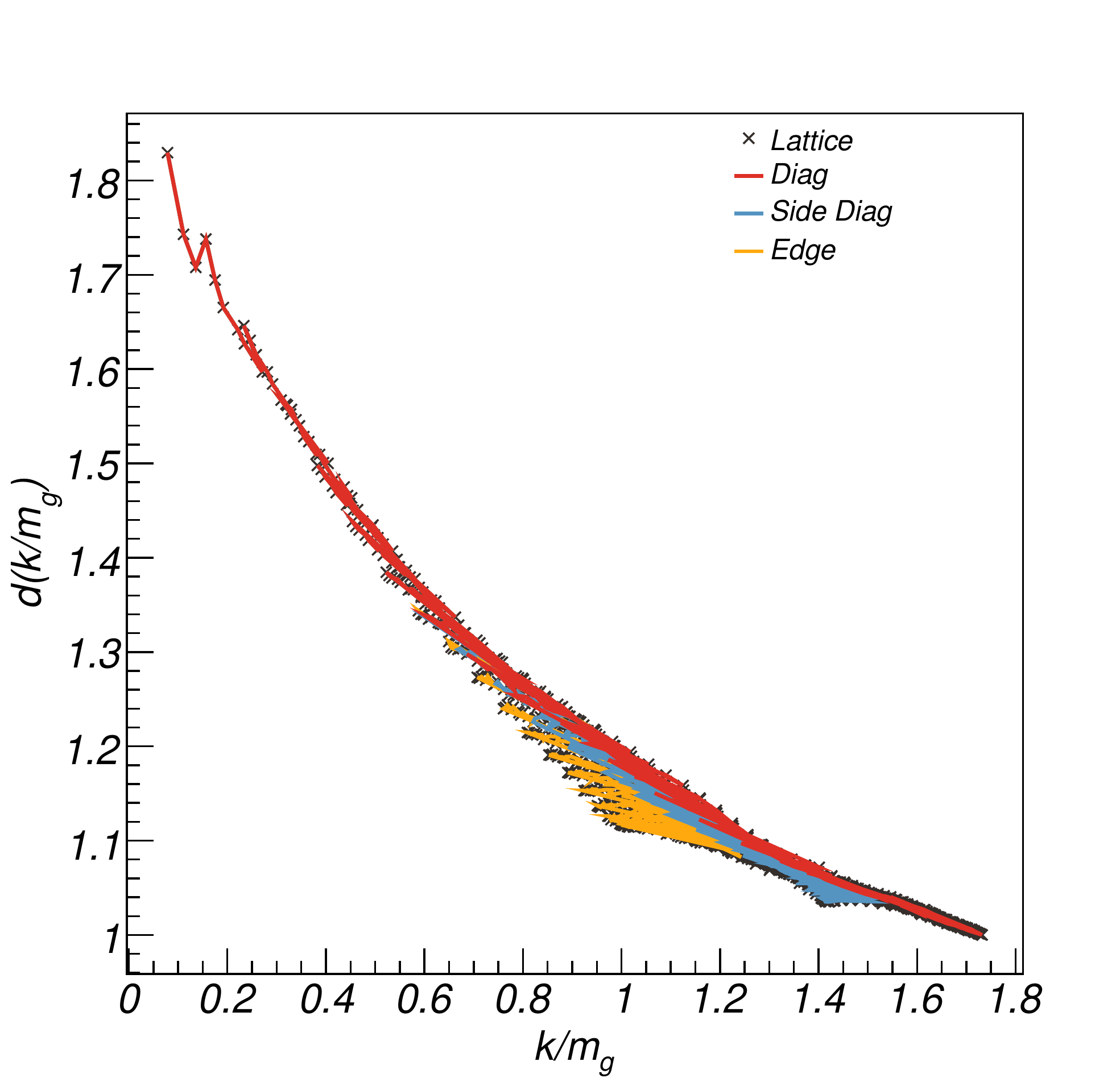}
}
\caption{(Color Online) Ghost dressing function for $d(k/m_{g})$ versus $k/m_{g}$ for $%
\hat m_{g}=1.25$ and $g=0.7$ on a lattice with $40^{3}$ volume. Here the crosses
denote the data points and the three lines connect the subsets of the points
laying within $0$ a) and $7$ b) minimum lattice momenta correspondingly of
the edge, the side diagonal and the diagonal directions of the momentum
lattice cube.}
\label{PLOT_GHOST_ROT_PROOF}
\end{figure}

It is argued in the Refs. \cite%
{Bonnet:2001uh,Leinweber:1998im,Marenzoni:1994ap} that using the redefined
lattice momentum variable of Eq.~(\ref{EQ_DEFINE_Q}) allows one to avoid the
leading-order discretization errors arising from the ultra-violet cutoff in
momentum introduced by the finite lattice spacing. Still, the errors from 
reducing the spatial rotational symmetry $O(3)$ group down to discrete $Z(3)$ 
are unaccounted for. These manifest themselves as a large spread in the 
ghost propagator. 
This spread occurs in a characteristic pattern, as can be seen in 
Fig.~\ref{PLOT_GHOST_DIAG}, which becomes more prominent with increased lattice
volume.  These patterns can be easily understood by considering a selection of 
subsets of the points plotted by using
certain criteria imposed on the momentum variable. The first subset
considered has the constraint that all three components of the momentum are
equal to each other (laying on the diagonal direction of the 
lattice).  This selection of the data forms a smooth line through the upper 
part of the plot. A subset including points with two of the momentum components
equal to each other and the third one set to zero (along the diagonal direction
of the cube's side) forms another smooth curve, this one going through the 
middle of the plot.  Finally, the subset with only one non-zero component of 
momentum (along the side of the cube) forms a line passing through the lowest 
part of the plot.  These subsets are shown in Fig.~\ref{PLOT_GHOST_ROT_PROOF}a.
Furthermore, if the constraints described above are allowed to be violated 
by a few units of minimum lattice momentum, the rest of the points in the plot 
start to fall into these subgroups, as shown in 
Fig.~\ref{PLOT_GHOST_ROT_PROOF}b.  Thus, for the further analysis of our data 
we will use only a subset of points with momentum components not differing from
each other by more than one unit of minimum lattice momentum.  This is  the 
\textquotedblleft cylinder cut\textquotedblright\ introduced in 
Refs.~\cite{Leinweber:1998im,Bonnet:2001uh}, which allows us to select the 
points least affected by errors introduced by the broken rotational symmetry 
and leaves a sufficient number of points for statistical analysis.

\subsubsection{Finite Volume Effects}

\begin{figure}[ptbh]
\begin{center}
\includegraphics[width=0.4\textwidth
]{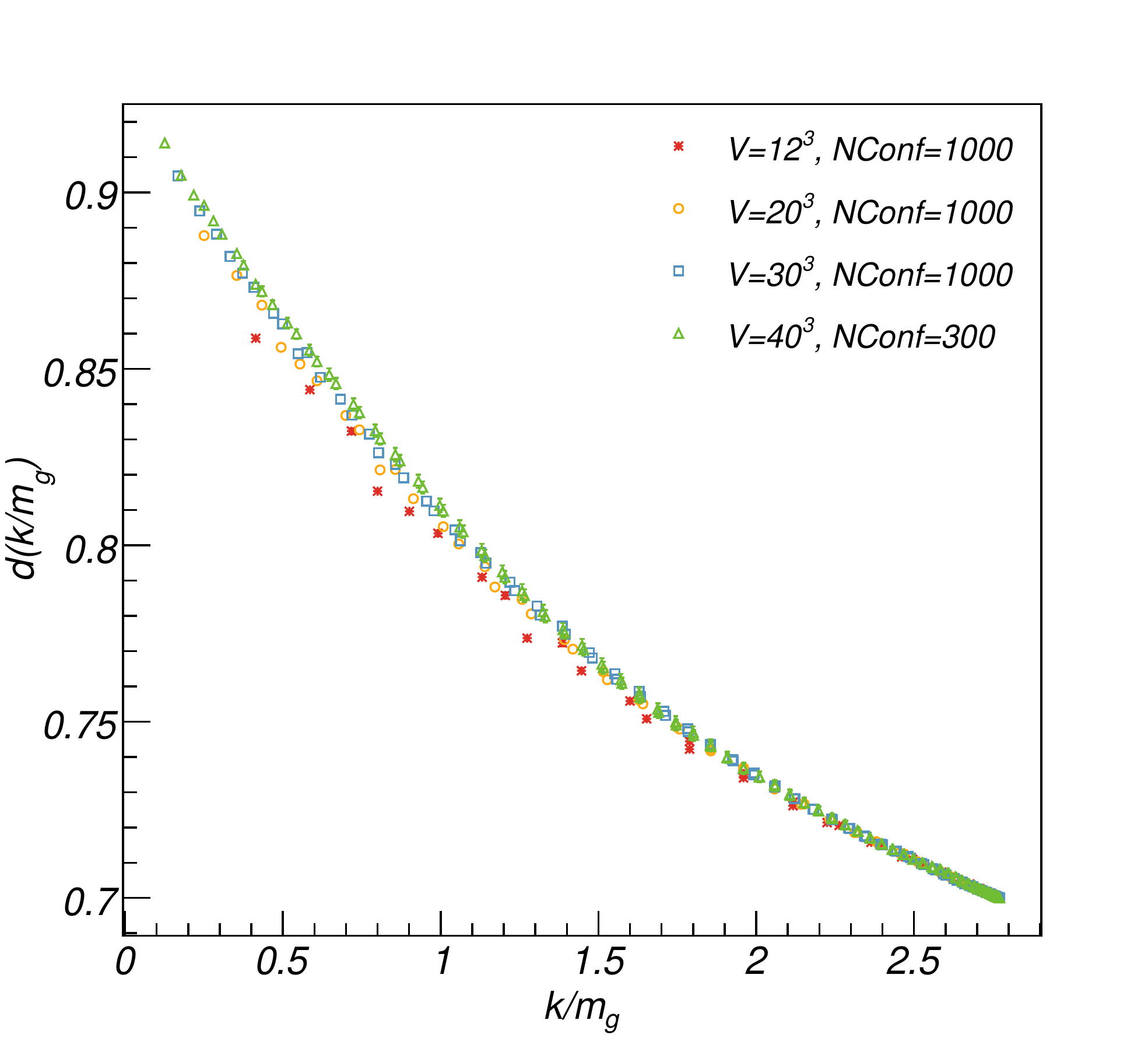}
\end{center}
\caption{(Color Online) Ghost dressing function $d(k/m_{g})$ versus $k/m_{g}$ for $%
\hat m_{g}=1.25$ and $g=0.7$. We show the influence of varying the lattice 
volume $V$ on the calculated data points with momentum components not differing
by more than two units lattice momentum, as described in the text. Here $NConf$
denotes the number of the sampled gluon configurations.}
\label{PLOT_GHOST_VOLUME}
\end{figure}

While a consistent treatment of the finite volume effects requires extensive
investigation into discretization of the theory on the lattice, here we
simply investigate this dependence by comparing benchmark calculations on
lattices with different volumes. A set of calculations with four different
lattice volumes is shown in Fig. \ref{PLOT_GHOST_VOLUME}, which shows that 
there are very small variations only in the low momenta region for lattice 
volumes from $20^{3}$ to $40^{3}$. 
Thus we choose to use lattice volume of $20^{3}$ for the further calculations, 
which allows for both reasonable computational time and small errors.

\subsection{Renormalization\label{SEC_RENORM}}

\begin{figure}[ptbh]
\begin{center}
\includegraphics[width=0.4\textwidth]{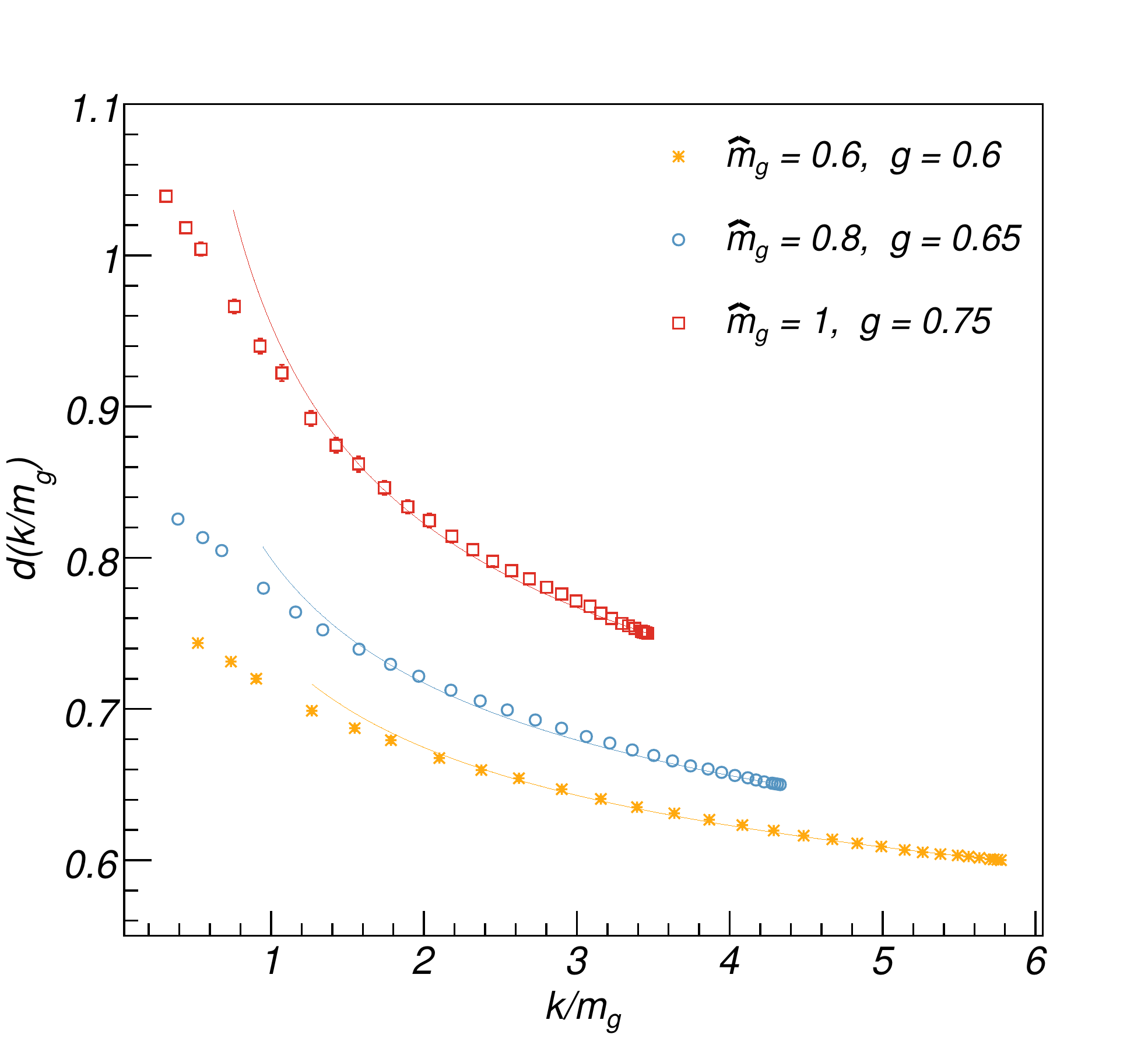}
\end{center}
\caption{(Color Online) Simultaneous fits to the ghost dressing function $d(k/m_{g})$
versus $k/m_{g}$ for $\hat m_{g}\gtrsim 2\protect\pi/N_{Lat}$.}
\label{PLOT_GHOST_LOW_MU}
\end{figure}

The introduction of a finite momentum grid provides a sharp cutoff for
the regularization of the ultra-violet divergences, {\it i.e.} it is 
equivalent to the role of $\Lambda$ in Eq.~(\ref{lambda}). 
In order to identify the ghost dressing function with the running coupling, 
for each lattice spacing it should be possible to choose the lattice 
coupling, $g$ in Eq.~(\ref{latwf}), so that the results of simulations are 
independent of the lattice spacing.  In the simulation, explicit dependence on 
the lattice spacing enters through dependence on $\hat m_g = a m_g$, {\it e.g.}
when the lattice ghost dressing function is plotted against 
$\tilde k = k/m_g = \hat k/\hat m_g$ the result should be independent of 
$\hat m_g$ and depend only on the value of the renormalized coupling.  In 
practice, we produce a series of simulations with different values of $g$ in 
the range between $g=0.5 - 1.5$ and $\hat m_g$ in the range of the accessible 
lattice momenta 
$2\pi/N_{Lat}\leq \hat m_{g} \leq \sqrt{3}\pi$.  We then compare with the 
scaling predicted by the solutions of the 
Dyson equations  given in Eqs.~(\ref{EQ_GHOST_REN_LR_LOW}) and
(\ref{EQ_GHOST_REN_LR_HIGH}), in the low and high momentum region, 
respectively. 
In the high momentum regime, $k/m_g= \hat k/\hat m_g$ is kept large by
running simulations with small $\hat m_g$ {\it i.e.} with $\hat m_{g}\gtrsim 2\pi/N_{Lat}$. In this regime the constituent gluon mass is close to the minimum 
accessible momentum scale on the lattice and the gluon propagator is 
close to asymptotic while the non-perturbative effects are only present for a 
few, lowest momentum points.  This regime should be 
described by Eq.~(\ref{EQ_GHOST_REN_LR_HIGH}), 
  \begin{eqnarray}
d( \tilde k) =\frac{d(\tilde \mu)}{\left[ 1+\beta'_H d^{1/\gamma'}(\tilde \mu )\log \left( \frac{\tilde k}{\tilde \mu}\right) \right]^{\gamma'}},
\end{eqnarray}%
where $\beta'_H$ and $\gamma'$ will be treated as fit parameters.
For $N_d=20$ each data set has $30$ momentum points after the imposed 
``diagonal'' cut described above.  We choose $6$ data sets 
with $\hat m_g \in [0.1,1]$  and $g \in [0.3, 0.75]$, where the $25$ highest 
momentum points can be considered to be in the asymptotic region.  For each 
value of the coupling, $g$, the value of $d(\tilde \mu)$ is fixed by the data 
itself with $\tilde \mu$ set equal to the momentum cut-off, $\tilde \mu = \sqrt{3}\pi/\hat m_g$. 
The formula in Eq.~(\ref{EQ_GHOST_REN_LR_HIGH}) is fitted to  all $150$  data points by varying  $\beta'_H$ and $\gamma'$. 
The resulting remarkably good fits are shown in 
Fig.~\ref{PLOT_GHOST_LOW_MU} with the best-fit value of $\beta'_H=0.86(2)$ and 
$\gamma'=0.5(2)$.  The data deviate from the perturbative form at intermediate 
momenta, which is to be expected.

\begin{figure}[ptb]
\begin{center}
\includegraphics[width=0.4\textwidth
]{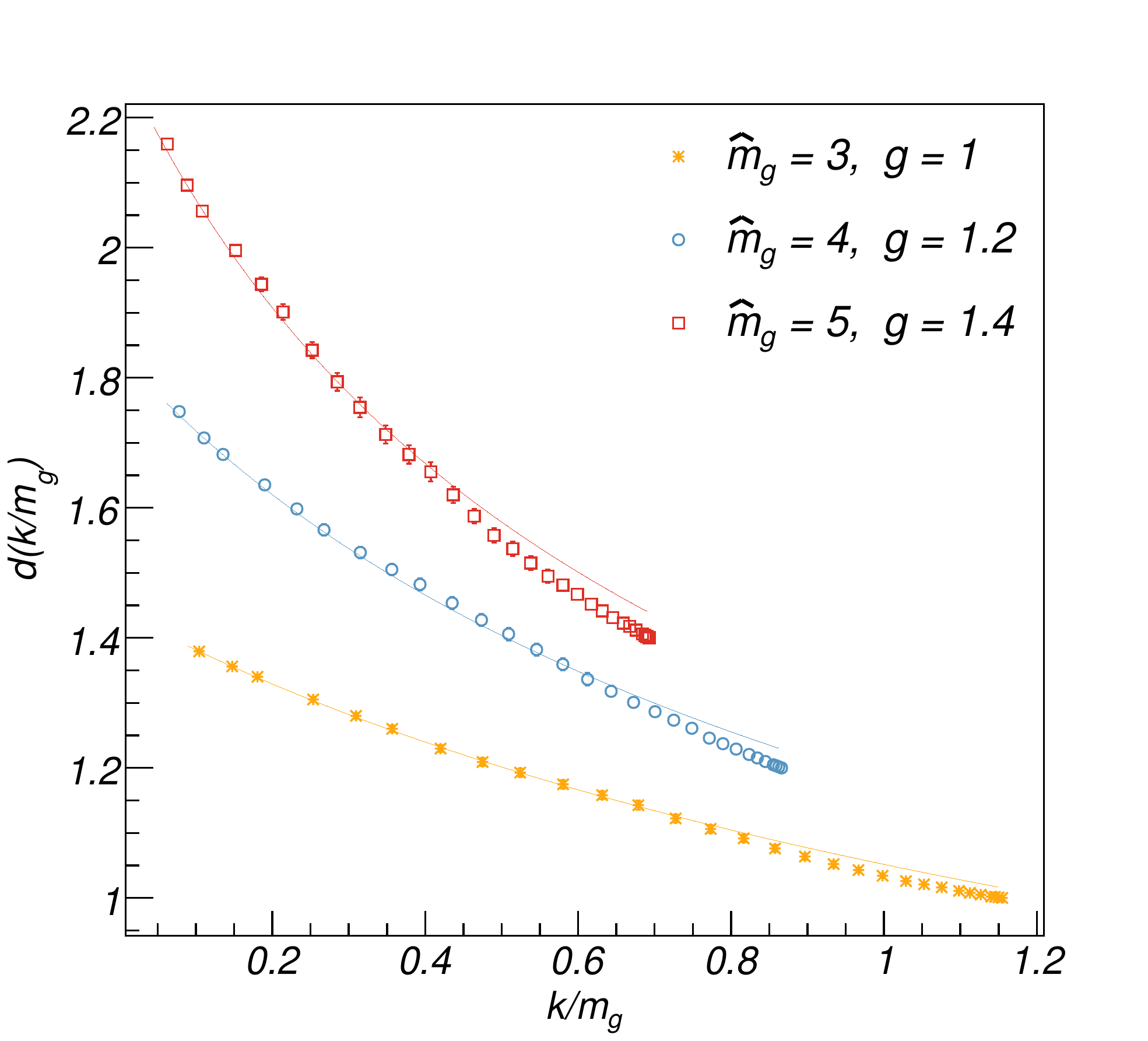}
\end{center}
\caption{(Color Online) Simultaneous fits to the ghost dressing function $d(k/m_{g})$
versus $k/m_{g}$ for $\hat m_{g}\lesssim\protect\sqrt{3}\protect\pi $.}
\label{PLOT_GHOST_HIGH_MU}
\end{figure}

On the other hand, in simulations with large $\hat m_{g}$ {\it i.e} for 
$\hat m_{g}\lesssim \sqrt{3}\pi$, Eq.~(\ref{EQ_GHOST_REN_LR_LOW}) should apply.
Then the constituent gluon mass is close to the largest accessible momentum 
scale on the lattice.  This regime is dominated by non-propagating gluons 
induced by non-perturbative dressing.  Here we expect,  
\begin{equation}
d(\tilde k) =\frac{d(\tilde \mu)}{\left[ 1+\beta'_L d^{1/\gamma'}(\tilde \mu )\left( \tilde k-\tilde \mu \right) \right]^{\gamma'}} .
\end{equation}%
In this region we select a total of $8$ data sets composed of 
$\hat m_g \in [3,5]$  and $g \in [1, 1.4]$, where the $15$ lowest momentum 
points can be considered to be in the non-perturbative region.  Here, 
$d(\tilde \mu)$ is obtained from each data set itself, at $\tilde \mu$ 
chosen, to avoid finite-volume effects, to be the second lowest momentum point.
In the low momentum range a total of $120$ data points was fitted varying 
$\beta'_L$ while keeping $\gamma'=1/2$ which was previously determined from the
high momentum fit.  A sample of data points with the corresponding fits are 
shown in Fig.~\ref{PLOT_GHOST_HIGH_MU} for the best fit value of
$\beta'_L=0.81(2)$.  Again, the discrepancies in the higher momentum region are
expected as a consequence of deviations from purely non-perturbative behavior 
set by the asymptotic tail of the gluon propagator.

\section{Conclusions}
\label{Summary} 

We have computed the ghost correlation function by direct Monte Carlo 
simulation of the functional integral with a model gaussian wave 
functional and compared it with the solution of the corresponding Dyson 
equation.  We have found that the scaling behavior
of the solution of the Dyson equation is reproduced in the simulation. This 
confirms that the corrections to the rainbow-ladder approximation 
are both IR and UV finite, and do not change the scaling properties. 
The $\beta$ function obtained from simulations is, however, an
order of magnitude larger than the one from the Dyson equation. This 
is to be expected, since the Dyson equation does not properly take 
into account the boundary of the field space integral, and thus is 
expected to overestimate the magnitude of the allowed field values and
thus of the critical coupling.  The Monte Carlo simulation still needs to have 
the Faddeev-Popov Jacobian implemented, but that is not expected to 
qualitatively change the results.

In our simulations we have found that positivity of the Faddeev-Popov operator 
is not sufficient to produce critical behavior.  This needs to be investigated 
further, in particular on larger volumes; nevertheless, since the simple 
gaussian vacuum wave functional does not probe topological configurations 
({\it  e.g.} of magnetic disorder) it is not too surprising that the IR 
enhancement of the ghost correlator or Coulomb form factor fails to 
quantitatively reproduce confinement. For this purpose a wave functional of the
type proposed in Ref.~\cite{Greensite:2007ij} should be tried.

\section{Acknowledgment}
We would like to thank  H.~Reinhardt  for  continuing discussions of the Coulomb gauge QCD.  This work was supported in part by the US Department of Energy grant under 
contract DE-FG0287ER40365.

\bibliographystyle{apsrev}
\bibliography{coulomb}

\begin{thebibliography}{61}
\expandafter\ifx\csname natexlab\endcsname\relax\def\natexlab#1{#1}\fi
\expandafter\ifx\csname bibnamefont\endcsname\relax
  \def\bibnamefont#1{#1}\fi
\expandafter\ifx\csname bibfnamefont\endcsname\relax
  \def\bibfnamefont#1{#1}\fi
\expandafter\ifx\csname citenamefont\endcsname\relax
  \def\citenamefont#1{#1}\fi
\expandafter\ifx\csname url\endcsname\relax
  \def\url#1{\texttt{#1}}\fi
\expandafter\ifx\csname urlprefix\endcsname\relax\def\urlprefix{URL }\fi
\providecommand{\bibinfo}[2]{#2}
\providecommand{\eprint}[2][]{\url{#2}}

\bibitem[{\citenamefont{Natale}(2007)}]{Natale:2006nv}
\bibinfo{author}{\bibfnamefont{A.~A.} \bibnamefont{Natale}},
  \bibinfo{journal}{Braz. J. Phys.} \textbf{\bibinfo{volume}{37}},
  \bibinfo{pages}{306} (\bibinfo{year}{2007}), \eprint{hep-ph/0610256}.

\bibitem[{\citenamefont{Fischer}(2006)}]{Fischer:2006ub}
\bibinfo{author}{\bibfnamefont{C.~S.} \bibnamefont{Fischer}},
  \bibinfo{journal}{J. Phys.} \textbf{\bibinfo{volume}{G32}},
  \bibinfo{pages}{R253} (\bibinfo{year}{2006}), \eprint{hep-ph/0605173}.

\bibitem[{\citenamefont{von Smekal et~al.}(1998)\citenamefont{von Smekal,
  Hauck, and Alkofer}}]{von_Smekal:1997vx}
\bibinfo{author}{\bibfnamefont{L.}~\bibnamefont{von Smekal}},
  \bibinfo{author}{\bibfnamefont{A.}~\bibnamefont{Hauck}}, \bibnamefont{and}
  \bibinfo{author}{\bibfnamefont{R.}~\bibnamefont{Alkofer}},
  \bibinfo{journal}{Ann. Phys.} \textbf{\bibinfo{volume}{267}},
  \bibinfo{pages}{1} (\bibinfo{year}{1998}), \eprint{hep-ph/9707327}.

\bibitem[{\citenamefont{Braun et~al.}(2007)\citenamefont{Braun, Gies, and
  Pawlowski}}]{Braun:2007bx}
\bibinfo{author}{\bibfnamefont{J.}~\bibnamefont{Braun}},
  \bibinfo{author}{\bibfnamefont{H.}~\bibnamefont{Gies}}, \bibnamefont{and}
  \bibinfo{author}{\bibfnamefont{J.~M.} \bibnamefont{Pawlowski}}
  (\bibinfo{year}{2007}), \eprint{0708.2413}.

\bibitem[{\citenamefont{Zwanziger}(2002)}]{Zwanziger:2001kw}
\bibinfo{author}{\bibfnamefont{D.}~\bibnamefont{Zwanziger}},
  \bibinfo{journal}{Phys. Rev.} \textbf{\bibinfo{volume}{D65}},
  \bibinfo{pages}{094039} (\bibinfo{year}{2002}), \eprint{hep-th/0109224}.

\bibitem[{\citenamefont{Pawlowski et~al.}(2004)\citenamefont{Pawlowski, Litim,
  Nedelko, and von Smekal}}]{Pawlowski:2003hq}
\bibinfo{author}{\bibfnamefont{J.~M.} \bibnamefont{Pawlowski}},
  \bibinfo{author}{\bibfnamefont{D.~F.} \bibnamefont{Litim}},
  \bibinfo{author}{\bibfnamefont{S.}~\bibnamefont{Nedelko}}, \bibnamefont{and}
  \bibinfo{author}{\bibfnamefont{L.}~\bibnamefont{von Smekal}},
  \bibinfo{journal}{Phys. Rev. Lett.} \textbf{\bibinfo{volume}{93}},
  \bibinfo{pages}{152002} (\bibinfo{year}{2004}), \eprint{hep-th/0312324}.

\bibitem[{\citenamefont{Lerche and von Smekal}(2002)}]{Lerche:2002ep}
\bibinfo{author}{\bibfnamefont{C.}~\bibnamefont{Lerche}} \bibnamefont{and}
  \bibinfo{author}{\bibfnamefont{L.}~\bibnamefont{von Smekal}},
  \bibinfo{journal}{Phys. Rev.} \textbf{\bibinfo{volume}{D65}},
  \bibinfo{pages}{125006} (\bibinfo{year}{2002}), \eprint{hep-ph/0202194}.

\bibitem[{\citenamefont{Aguilar and Natale}(2004)}]{Aguilar:2004sw}
\bibinfo{author}{\bibfnamefont{A.~C.} \bibnamefont{Aguilar}} \bibnamefont{and}
  \bibinfo{author}{\bibfnamefont{A.~A.} \bibnamefont{Natale}},
  \bibinfo{journal}{JHEP} \textbf{\bibinfo{volume}{08}}, \bibinfo{pages}{057}
  (\bibinfo{year}{2004}), \eprint{hep-ph/0408254}.

\bibitem[{\citenamefont{Frasca}(2007)}]{Frasca:2007uz}
\bibinfo{author}{\bibfnamefont{M.}~\bibnamefont{Frasca}}
  (\bibinfo{year}{2007}), \eprint{0709.2042}.

\bibitem[{\citenamefont{Boucaud et~al.}(2007{\natexlab{a}})}]{Boucaud:2007va}
\bibinfo{author}{\bibfnamefont{P.}~\bibnamefont{Boucaud}} \bibnamefont{et~al.},
  \bibinfo{journal}{Eur. Phys. J.} \textbf{\bibinfo{volume}{A31}},
  \bibinfo{pages}{750} (\bibinfo{year}{2007}{\natexlab{a}}),
  \eprint{hep-ph/0701114}.

\bibitem[{\citenamefont{Boucaud et~al.}(2007{\natexlab{b}})}]{Boucaud:2007hy}
\bibinfo{author}{\bibfnamefont{P.}~\bibnamefont{Boucaud}} \bibnamefont{et~al.},
  \bibinfo{journal}{JHEP} \textbf{\bibinfo{volume}{03}}, \bibinfo{pages}{076}
  (\bibinfo{year}{2007}{\natexlab{b}}), \eprint{hep-ph/0702092}.

\bibitem[{\citenamefont{Chernodub and Zakharov}(2007)}]{Chernodub:2007rn}
\bibinfo{author}{\bibfnamefont{M.~N.} \bibnamefont{Chernodub}}
  \bibnamefont{and} \bibinfo{author}{\bibfnamefont{V.~I.}
  \bibnamefont{Zakharov}} (\bibinfo{year}{2007}), \eprint{hep-ph/0703167}.

\bibitem[{\citenamefont{Dudal et~al.}(2007)\citenamefont{Dudal, Sorella,
  Vandersickel, and Verschelde}}]{Dudal:2007cw}
\bibinfo{author}{\bibfnamefont{D.}~\bibnamefont{Dudal}},
  \bibinfo{author}{\bibfnamefont{S.~P.} \bibnamefont{Sorella}},
  \bibinfo{author}{\bibfnamefont{N.}~\bibnamefont{Vandersickel}},
  \bibnamefont{and}
  \bibinfo{author}{\bibfnamefont{H.}~\bibnamefont{Verschelde}}
  (\bibinfo{year}{2007}), \eprint{0711.4496}.

\bibitem[{\citenamefont{Cucchieri}(1997)}]{Cucchieri:1997dx}
\bibinfo{author}{\bibfnamefont{A.}~\bibnamefont{Cucchieri}},
  \bibinfo{journal}{Nucl. Phys.} \textbf{\bibinfo{volume}{B508}},
  \bibinfo{pages}{353} (\bibinfo{year}{1997}), \eprint{hep-lat/9705005}.

\bibitem[{\citenamefont{Ilgenfritz et~al.}(2007)\citenamefont{Ilgenfritz,
  Muller-Preussker, Sternbeck, Schiller, and Bogolubsky}}]{Ilgenfritz:2006he}
\bibinfo{author}{\bibfnamefont{E.~M.} \bibnamefont{Ilgenfritz}},
  \bibinfo{author}{\bibfnamefont{M.}~\bibnamefont{Muller-Preussker}},
  \bibinfo{author}{\bibfnamefont{A.}~\bibnamefont{Sternbeck}},
  \bibinfo{author}{\bibfnamefont{A.}~\bibnamefont{Schiller}}, \bibnamefont{and}
  \bibinfo{author}{\bibfnamefont{I.~L.} \bibnamefont{Bogolubsky}},
  \bibinfo{journal}{Braz. J. Phys.} \textbf{\bibinfo{volume}{37}},
  \bibinfo{pages}{193} (\bibinfo{year}{2007}), \eprint{hep-lat/0609043}.

\bibitem[{\citenamefont{Leinweber et~al.}(1998)\citenamefont{Leinweber,
  Skullerud, Williams, and Parrinello}}]{Leinweber:1998im}
\bibinfo{author}{\bibfnamefont{D.~B.} \bibnamefont{Leinweber}},
  \bibinfo{author}{\bibfnamefont{J.~I.} \bibnamefont{Skullerud}},
  \bibinfo{author}{\bibfnamefont{A.~G.} \bibnamefont{Williams}},
  \bibnamefont{and}
  \bibinfo{author}{\bibfnamefont{C.}~\bibnamefont{Parrinello}}
  (\bibinfo{collaboration}{UKQCD}), \bibinfo{journal}{Phys. Rev.}
  \textbf{\bibinfo{volume}{D58}}, \bibinfo{pages}{031501}
  (\bibinfo{year}{1998}), \eprint{hep-lat/9803015}.

\bibitem[{\citenamefont{Furui and Nakajima}(2004)}]{Furui:2003jr}
\bibinfo{author}{\bibfnamefont{S.}~\bibnamefont{Furui}} \bibnamefont{and}
  \bibinfo{author}{\bibfnamefont{H.}~\bibnamefont{Nakajima}},
  \bibinfo{journal}{Phys. Rev.} \textbf{\bibinfo{volume}{D69}},
  \bibinfo{pages}{074505} (\bibinfo{year}{2004}), \eprint{hep-lat/0305010}.

\bibitem[{\citenamefont{Cucchieri and
  Mendes}(2007{\natexlab{a}})}]{Cucchieri:2006xi}
\bibinfo{author}{\bibfnamefont{A.}~\bibnamefont{Cucchieri}} \bibnamefont{and}
  \bibinfo{author}{\bibfnamefont{T.}~\bibnamefont{Mendes}},
  \bibinfo{journal}{Braz. J. Phys.} \textbf{\bibinfo{volume}{37}},
  \bibinfo{pages}{484} (\bibinfo{year}{2007}{\natexlab{a}}),
  \eprint{hep-ph/0605224}.

\bibitem[{\citenamefont{Cucchieri}(1999)}]{Cucchieri:1999sz}
\bibinfo{author}{\bibfnamefont{A.}~\bibnamefont{Cucchieri}},
  \bibinfo{journal}{Phys. Rev.} \textbf{\bibinfo{volume}{D60}},
  \bibinfo{pages}{034508} (\bibinfo{year}{1999}), \eprint{hep-lat/9902023}.

\bibitem[{\citenamefont{Cucchieri et~al.}(2003)\citenamefont{Cucchieri, Mendes,
  and Taurines}}]{Cucchieri:2003di}
\bibinfo{author}{\bibfnamefont{A.}~\bibnamefont{Cucchieri}},
  \bibinfo{author}{\bibfnamefont{T.}~\bibnamefont{Mendes}}, \bibnamefont{and}
  \bibinfo{author}{\bibfnamefont{A.~R.} \bibnamefont{Taurines}},
  \bibinfo{journal}{Phys. Rev.} \textbf{\bibinfo{volume}{D67}},
  \bibinfo{pages}{091502} (\bibinfo{year}{2003}), \eprint{hep-lat/0302022}.

\bibitem[{\citenamefont{Sternbeck et~al.}(2005)\citenamefont{Sternbeck,
  Ilgenfritz, Mueller-Preussker, and Schiller}}]{Sternbeck:2005tk}
\bibinfo{author}{\bibfnamefont{A.}~\bibnamefont{Sternbeck}},
  \bibinfo{author}{\bibfnamefont{E.~M.} \bibnamefont{Ilgenfritz}},
  \bibinfo{author}{\bibfnamefont{M.}~\bibnamefont{Mueller-Preussker}},
  \bibnamefont{and} \bibinfo{author}{\bibfnamefont{A.}~\bibnamefont{Schiller}},
  \bibinfo{journal}{Phys. Rev.} \textbf{\bibinfo{volume}{D72}},
  \bibinfo{pages}{014507} (\bibinfo{year}{2005}), \eprint{hep-lat/0506007}.

\bibitem[{\citenamefont{Boucaud et~al.}(2005)}]{Boucaud:2005gg}
\bibinfo{author}{\bibfnamefont{P.}~\bibnamefont{Boucaud}} \bibnamefont{et~al.},
  \bibinfo{journal}{Phys. Rev.} \textbf{\bibinfo{volume}{D72}},
  \bibinfo{pages}{114503} (\bibinfo{year}{2005}), \eprint{hep-lat/0506031}.

\bibitem[{\citenamefont{Bogolubsky et~al.}(2006)\citenamefont{Bogolubsky,
  Burgio, Muller-Preussker, and Mitrjushkin}}]{Bogolubsky:2005wf}
\bibinfo{author}{\bibfnamefont{I.~L.} \bibnamefont{Bogolubsky}},
  \bibinfo{author}{\bibfnamefont{G.}~\bibnamefont{Burgio}},
  \bibinfo{author}{\bibfnamefont{M.}~\bibnamefont{Muller-Preussker}},
  \bibnamefont{and} \bibinfo{author}{\bibfnamefont{V.~K.}
  \bibnamefont{Mitrjushkin}}, \bibinfo{journal}{Phys. Rev.}
  \textbf{\bibinfo{volume}{D74}}, \bibinfo{pages}{034503}
  (\bibinfo{year}{2006}), \eprint{hep-lat/0511056}.

\bibitem[{\citenamefont{Cucchieri et~al.}(2006)\citenamefont{Cucchieri, Maas,
  and Mendes}}]{Cucchieri:2006tf}
\bibinfo{author}{\bibfnamefont{A.}~\bibnamefont{Cucchieri}},
  \bibinfo{author}{\bibfnamefont{A.}~\bibnamefont{Maas}}, \bibnamefont{and}
  \bibinfo{author}{\bibfnamefont{T.}~\bibnamefont{Mendes}},
  \bibinfo{journal}{Phys. Rev.} \textbf{\bibinfo{volume}{D74}},
  \bibinfo{pages}{014503} (\bibinfo{year}{2006}), \eprint{hep-lat/0605011}.

\bibitem[{\citenamefont{Oliveira and
  Silva}(2007{\natexlab{a}})}]{Oliveira:2006zg}
\bibinfo{author}{\bibfnamefont{O.}~\bibnamefont{Oliveira}} \bibnamefont{and}
  \bibinfo{author}{\bibfnamefont{P.~J.} \bibnamefont{Silva}},
  \bibinfo{journal}{Braz. J. Phys.} \textbf{\bibinfo{volume}{37}},
  \bibinfo{pages}{201} (\bibinfo{year}{2007}{\natexlab{a}}),
  \eprint{hep-lat/0609036}.

\bibitem[{\citenamefont{Oliveira and
  Silva}(2007{\natexlab{b}})}]{Oliveira:2006yw}
\bibinfo{author}{\bibfnamefont{O.}~\bibnamefont{Oliveira}} \bibnamefont{and}
  \bibinfo{author}{\bibfnamefont{P.~J.} \bibnamefont{Silva}},
  \bibinfo{journal}{Eur. Phys. J.} \textbf{\bibinfo{volume}{A31}},
  \bibinfo{pages}{790} (\bibinfo{year}{2007}{\natexlab{b}}),
  \eprint{hep-lat/0609027}.

\bibitem[{\citenamefont{Fischer et~al.}(2002)\citenamefont{Fischer, Alkofer,
  and Reinhardt}}]{Fischer:2002eq}
\bibinfo{author}{\bibfnamefont{C.~S.} \bibnamefont{Fischer}},
  \bibinfo{author}{\bibfnamefont{R.}~\bibnamefont{Alkofer}}, \bibnamefont{and}
  \bibinfo{author}{\bibfnamefont{H.}~\bibnamefont{Reinhardt}},
  \bibinfo{journal}{Phys. Rev.} \textbf{\bibinfo{volume}{D65}},
  \bibinfo{pages}{094008} (\bibinfo{year}{2002}), \eprint{hep-ph/0202195}.

\bibitem[{\citenamefont{Fischer et~al.}(2006)\citenamefont{Fischer, Gruter, and
  Alkofer}}]{Fischer:2005ui}
\bibinfo{author}{\bibfnamefont{C.~S.} \bibnamefont{Fischer}},
  \bibinfo{author}{\bibfnamefont{B.}~\bibnamefont{Gruter}}, \bibnamefont{and}
  \bibinfo{author}{\bibfnamefont{R.}~\bibnamefont{Alkofer}},
  \bibinfo{journal}{Ann. Phys.} \textbf{\bibinfo{volume}{321}},
  \bibinfo{pages}{1918} (\bibinfo{year}{2006}), \eprint{hep-ph/0506053}.

\bibitem[{\citenamefont{Fischer
  et~al.}(2007{\natexlab{a}})\citenamefont{Fischer, Maas, Pawlowski, and von
  Smekal}}]{Fischer:2007pf}
\bibinfo{author}{\bibfnamefont{C.~S.} \bibnamefont{Fischer}},
  \bibinfo{author}{\bibfnamefont{A.}~\bibnamefont{Maas}},
  \bibinfo{author}{\bibfnamefont{J.~M.} \bibnamefont{Pawlowski}},
  \bibnamefont{and} \bibinfo{author}{\bibfnamefont{L.}~\bibnamefont{von
  Smekal}}, \bibinfo{journal}{Annals Phys.} \textbf{\bibinfo{volume}{322}},
  \bibinfo{pages}{2916} (\bibinfo{year}{2007}{\natexlab{a}}),
  \eprint{hep-ph/0701050}.

\bibitem[{\citenamefont{Fischer
  et~al.}(2007{\natexlab{b}})\citenamefont{Fischer, Alkofer, Maas, Pawlowski,
  and von Smekal}}]{Fischer:2007mc}
\bibinfo{author}{\bibfnamefont{C.~S.} \bibnamefont{Fischer}},
  \bibinfo{author}{\bibfnamefont{R.}~\bibnamefont{Alkofer}},
  \bibinfo{author}{\bibfnamefont{A.}~\bibnamefont{Maas}},
  \bibinfo{author}{\bibfnamefont{J.~M.} \bibnamefont{Pawlowski}},
  \bibnamefont{and} \bibinfo{author}{\bibfnamefont{L.}~\bibnamefont{von
  Smekal}}, \bibinfo{journal}{POS} \textbf{\bibinfo{volume}{LAT2007}},
  \bibinfo{pages}{300} (\bibinfo{year}{2007}{\natexlab{b}}),
  \eprint{0709.3205}.

\bibitem[{\citenamefont{Cucchieri and
  Mendes}(2007{\natexlab{b}})}]{Cucchieri:2007md}
\bibinfo{author}{\bibfnamefont{A.}~\bibnamefont{Cucchieri}} \bibnamefont{and}
  \bibinfo{author}{\bibfnamefont{T.}~\bibnamefont{Mendes}}
  (\bibinfo{year}{2007}{\natexlab{b}}), \eprint{0710.0412}.

\bibitem[{\citenamefont{Bogolubsky et~al.}(2007)\citenamefont{Bogolubsky,
  Ilgenfritz, Muller-Preussker, and Sternbeck}}]{Bogolubsky:2007ud}
\bibinfo{author}{\bibfnamefont{I.~L.} \bibnamefont{Bogolubsky}},
  \bibinfo{author}{\bibfnamefont{E.~M.} \bibnamefont{Ilgenfritz}},
  \bibinfo{author}{\bibfnamefont{M.}~\bibnamefont{Muller-Preussker}},
  \bibnamefont{and} \bibinfo{author}{\bibfnamefont{A.}~\bibnamefont{Sternbeck}}
  (\bibinfo{year}{2007}), \eprint{0710.1968}.

\bibitem[{\citenamefont{Sternbeck et~al.}(2007)\citenamefont{Sternbeck, von
  Smekal, Leinweber, and Williams}}]{Sternbeck:2007ug}
\bibinfo{author}{\bibfnamefont{A.}~\bibnamefont{Sternbeck}},
  \bibinfo{author}{\bibfnamefont{L.}~\bibnamefont{von Smekal}},
  \bibinfo{author}{\bibfnamefont{D.~B.} \bibnamefont{Leinweber}},
  \bibnamefont{and} \bibinfo{author}{\bibfnamefont{A.~G.}
  \bibnamefont{Williams}} (\bibinfo{year}{2007}), \eprint{0710.1982}.

\bibitem[{\citenamefont{Cucchieri et~al.}(2007)\citenamefont{Cucchieri, Mendes,
  Oliveira, and Silva}}]{Cucchieri:2007zm}
\bibinfo{author}{\bibfnamefont{A.}~\bibnamefont{Cucchieri}},
  \bibinfo{author}{\bibfnamefont{T.}~\bibnamefont{Mendes}},
  \bibinfo{author}{\bibfnamefont{O.}~\bibnamefont{Oliveira}}, \bibnamefont{and}
  \bibinfo{author}{\bibfnamefont{P.~J.} \bibnamefont{Silva}},
  \bibinfo{journal}{Phys. Rev.} \textbf{\bibinfo{volume}{D76}},
  \bibinfo{pages}{114507} (\bibinfo{year}{2007}), \eprint{0705.3367}.

\bibitem[{\citenamefont{Cucchieri and
  Mendes}(2007{\natexlab{c}})}]{Cucchieri:2007rg}
\bibinfo{author}{\bibfnamefont{A.}~\bibnamefont{Cucchieri}} \bibnamefont{and}
  \bibinfo{author}{\bibfnamefont{T.}~\bibnamefont{Mendes}}
  (\bibinfo{year}{2007}{\natexlab{c}}), \eprint{0712.3517}.

\bibitem[{\citenamefont{Bowman et~al.}(2007)}]{Bowman:2007du}
\bibinfo{author}{\bibfnamefont{P.~O.} \bibnamefont{Bowman}}
  \bibnamefont{et~al.}, \bibinfo{journal}{Phys. Rev.}
  \textbf{\bibinfo{volume}{D76}}, \bibinfo{pages}{094505}
  (\bibinfo{year}{2007}), \eprint{hep-lat/0703022}.

\bibitem[{\citenamefont{Kugo and Ojima}(1979)}]{Kugo:1979gm}
\bibinfo{author}{\bibfnamefont{T.}~\bibnamefont{Kugo}} \bibnamefont{and}
  \bibinfo{author}{\bibfnamefont{I.}~\bibnamefont{Ojima}},
  \bibinfo{journal}{Prog. Theor. Phys. Suppl.} \textbf{\bibinfo{volume}{66}},
  \bibinfo{pages}{1} (\bibinfo{year}{1979}).

\bibitem[{\citenamefont{Kugo}(1995)}]{Kugo:1995km}
\bibinfo{author}{\bibfnamefont{T.}~\bibnamefont{Kugo}} (\bibinfo{year}{1995}),
  \eprint{hep-th/9511033}.

\bibitem[{\citenamefont{Caudy and Greensite}(2007)}]{Caudy:2007sf}
\bibinfo{author}{\bibfnamefont{W.}~\bibnamefont{Caudy}} \bibnamefont{and}
  \bibinfo{author}{\bibfnamefont{J.}~\bibnamefont{Greensite}}
  (\bibinfo{year}{2007}), \eprint{0712.0999}.

\bibitem[{\citenamefont{Zwanziger}(2003)}]{Zwanziger:2002sh}
\bibinfo{author}{\bibfnamefont{D.}~\bibnamefont{Zwanziger}},
  \bibinfo{journal}{Phys. Rev. Lett.} \textbf{\bibinfo{volume}{90}},
  \bibinfo{pages}{102001} (\bibinfo{year}{2003}), \eprint{hep-lat/0209105}.

\bibitem[{\citenamefont{Zwanziger}(1997)}]{Zwanziger:1995cv}
\bibinfo{author}{\bibfnamefont{D.}~\bibnamefont{Zwanziger}},
  \bibinfo{journal}{Nucl. Phys.} \textbf{\bibinfo{volume}{B485}},
  \bibinfo{pages}{185} (\bibinfo{year}{1997}), \eprint{hep-th/9603203}.

\bibitem[{\citenamefont{Epple et~al.}(2007{\natexlab{a}})\citenamefont{Epple,
  Reinhardt, and Schleifenbaum}}]{Epple:2006hv}
\bibinfo{author}{\bibfnamefont{D.}~\bibnamefont{Epple}},
  \bibinfo{author}{\bibfnamefont{H.}~\bibnamefont{Reinhardt}},
  \bibnamefont{and}
  \bibinfo{author}{\bibfnamefont{W.}~\bibnamefont{Schleifenbaum}},
  \bibinfo{journal}{Phys. Rev.} \textbf{\bibinfo{volume}{D75}},
  \bibinfo{pages}{045011} (\bibinfo{year}{2007}{\natexlab{a}}),
  \eprint{hep-th/0612241}.

\bibitem[{\citenamefont{Cucchieri}(2007)}]{Cucchieri:2006hi}
\bibinfo{author}{\bibfnamefont{A.}~\bibnamefont{Cucchieri}},
  \bibinfo{journal}{AIP Conf. Proc.} \textbf{\bibinfo{volume}{892}},
  \bibinfo{pages}{22} (\bibinfo{year}{2007}), \eprint{hep-lat/0612004}.

\bibitem[{\citenamefont{Cucchieri and Zwanziger}(1997)}]{Cucchieri:1996ja}
\bibinfo{author}{\bibfnamefont{A.}~\bibnamefont{Cucchieri}} \bibnamefont{and}
  \bibinfo{author}{\bibfnamefont{D.}~\bibnamefont{Zwanziger}},
  \bibinfo{journal}{Phys. Rev. Lett.} \textbf{\bibinfo{volume}{78}},
  \bibinfo{pages}{3814} (\bibinfo{year}{1997}), \eprint{hep-th/9607224}.

\bibitem[{\citenamefont{Epple et~al.}(2007{\natexlab{b}})\citenamefont{Epple,
  Reinhardt, Schleifenbaum, and Szczepaniak}}]{Epple:2007ut}
\bibinfo{author}{\bibfnamefont{D.}~\bibnamefont{Epple}},
  \bibinfo{author}{\bibfnamefont{H.}~\bibnamefont{Reinhardt}},
  \bibinfo{author}{\bibfnamefont{W.}~\bibnamefont{Schleifenbaum}},
  \bibnamefont{and} \bibinfo{author}{\bibfnamefont{A.~P.}
  \bibnamefont{Szczepaniak}} (\bibinfo{year}{2007}{\natexlab{b}}),
  \eprint{0712.3694}.

\bibitem[{\citenamefont{Gribov}(1978)}]{Gribov:1977wm}
\bibinfo{author}{\bibfnamefont{V.~N.} \bibnamefont{Gribov}},
  \bibinfo{journal}{Nucl. Phys.} \textbf{\bibinfo{volume}{B139}},
  \bibinfo{pages}{1} (\bibinfo{year}{1978}).

\bibitem[{\citenamefont{Zwanziger}(1994)}]{Zwanziger:1993dh}
\bibinfo{author}{\bibfnamefont{D.}~\bibnamefont{Zwanziger}},
  \bibinfo{journal}{Nucl. Phys.} \textbf{\bibinfo{volume}{B412}},
  \bibinfo{pages}{657} (\bibinfo{year}{1994}).

\bibitem[{\citenamefont{Dokshitzer and Kharzeev}(2004)}]{Dokshitzer:2004ie}
\bibinfo{author}{\bibfnamefont{Y.~L.} \bibnamefont{Dokshitzer}}
  \bibnamefont{and} \bibinfo{author}{\bibfnamefont{D.~E.}
  \bibnamefont{Kharzeev}}, \bibinfo{journal}{Ann. Rev. Nucl. Part. Sci.}
  \textbf{\bibinfo{volume}{54}}, \bibinfo{pages}{487} (\bibinfo{year}{2004}),
  \eprint{hep-ph/0404216}.

\bibitem[{\citenamefont{Zwanziger}(1991{\natexlab{a}})}]{Zwanziger:1990by}
\bibinfo{author}{\bibfnamefont{D.}~\bibnamefont{Zwanziger}},
  \bibinfo{journal}{Phys. Lett.} \textbf{\bibinfo{volume}{B257}},
  \bibinfo{pages}{168} (\bibinfo{year}{1991}{\natexlab{a}}).

\bibitem[{\citenamefont{Zwanziger}(1991{\natexlab{b}})}]{Zwanziger:1991gz}
\bibinfo{author}{\bibfnamefont{D.}~\bibnamefont{Zwanziger}},
  \bibinfo{journal}{Nucl. Phys.} \textbf{\bibinfo{volume}{B364}},
  \bibinfo{pages}{127} (\bibinfo{year}{1991}{\natexlab{b}}).

\bibitem[{\citenamefont{Szczepaniak and Swanson}(2002)}]{Szczepaniak:2001rg}
\bibinfo{author}{\bibfnamefont{A.~P.} \bibnamefont{Szczepaniak}}
  \bibnamefont{and} \bibinfo{author}{\bibfnamefont{E.~S.}
  \bibnamefont{Swanson}}, \bibinfo{journal}{Phys. Rev.}
  \textbf{\bibinfo{volume}{D65}}, \bibinfo{pages}{025012}
  (\bibinfo{year}{2002}), \eprint{hep-ph/0107078}.

\bibitem[{\citenamefont{Szczepaniak}(2004)}]{Szczepaniak:2003ve}
\bibinfo{author}{\bibfnamefont{A.~P.} \bibnamefont{Szczepaniak}},
  \bibinfo{journal}{Phys. Rev.} \textbf{\bibinfo{volume}{D69}},
  \bibinfo{pages}{074031} (\bibinfo{year}{2004}), \eprint{hep-ph/0306030}.

\bibitem[{\citenamefont{Feuchter and
  Reinhardt}(2004{\natexlab{a}})}]{Feuchter:2004mk}
\bibinfo{author}{\bibfnamefont{C.}~\bibnamefont{Feuchter}} \bibnamefont{and}
  \bibinfo{author}{\bibfnamefont{H.}~\bibnamefont{Reinhardt}},
  \bibinfo{journal}{Phys. Rev.} \textbf{\bibinfo{volume}{D70}},
  \bibinfo{pages}{105021} (\bibinfo{year}{2004}{\natexlab{a}}),
  \eprint{hep-th/0408236}.

\bibitem[{\citenamefont{Feuchter and
  Reinhardt}(2004{\natexlab{b}})}]{Feuchter:2004gb}
\bibinfo{author}{\bibfnamefont{C.}~\bibnamefont{Feuchter}} \bibnamefont{and}
  \bibinfo{author}{\bibfnamefont{H.}~\bibnamefont{Reinhardt}}
  (\bibinfo{year}{2004}{\natexlab{b}}), \eprint{hep-th/0402106}.

\bibitem[{\citenamefont{Reinhardt and Feuchter}(2005)}]{Reinhardt:2004mm}
\bibinfo{author}{\bibfnamefont{H.}~\bibnamefont{Reinhardt}} \bibnamefont{and}
  \bibinfo{author}{\bibfnamefont{C.}~\bibnamefont{Feuchter}},
  \bibinfo{journal}{Phys. Rev.} \textbf{\bibinfo{volume}{D71}},
  \bibinfo{pages}{105002} (\bibinfo{year}{2005}), \eprint{hep-th/0408237}.

\bibitem[{\citenamefont{Christ and Lee}(1980)}]{Christ:1980ku}
\bibinfo{author}{\bibfnamefont{N.~H.} \bibnamefont{Christ}} \bibnamefont{and}
  \bibinfo{author}{\bibfnamefont{T.~D.} \bibnamefont{Lee}},
  \bibinfo{journal}{Phys. Rev.} \textbf{\bibinfo{volume}{D22}},
  \bibinfo{pages}{939} (\bibinfo{year}{1980}).

\bibitem[{\citenamefont{Zwanziger}(2004)}]{Zwanziger:2003cf}
\bibinfo{author}{\bibfnamefont{D.}~\bibnamefont{Zwanziger}},
  \bibinfo{journal}{Phys. Rev.} \textbf{\bibinfo{volume}{D69}},
  \bibinfo{pages}{016002} (\bibinfo{year}{2004}), \eprint{hep-ph/0303028}.

\bibitem[{\citenamefont{Swift}(1988)}]{Swift:1988za}
\bibinfo{author}{\bibfnamefont{A.~R.} \bibnamefont{Swift}},
  \bibinfo{journal}{Phys. Rev.} \textbf{\bibinfo{volume}{D38}},
  \bibinfo{pages}{668} (\bibinfo{year}{1988}).

\bibitem[{\citenamefont{Bonnet et~al.}(2001)\citenamefont{Bonnet, Bowman,
  Leinweber, Williams, and Zanotti}}]{Bonnet:2001uh}
\bibinfo{author}{\bibfnamefont{F.~D.~R.} \bibnamefont{Bonnet}},
  \bibinfo{author}{\bibfnamefont{P.~O.} \bibnamefont{Bowman}},
  \bibinfo{author}{\bibfnamefont{D.~B.} \bibnamefont{Leinweber}},
  \bibinfo{author}{\bibfnamefont{A.~G.} \bibnamefont{Williams}},
  \bibnamefont{and} \bibinfo{author}{\bibfnamefont{J.~M.}
  \bibnamefont{Zanotti}}, \bibinfo{journal}{Phys. Rev.}
  \textbf{\bibinfo{volume}{D64}}, \bibinfo{pages}{034501}
  (\bibinfo{year}{2001}), \eprint{hep-lat/0101013}.

\bibitem[{\citenamefont{Marenzoni et~al.}(1995)\citenamefont{Marenzoni,
  Martinelli, and Stella}}]{Marenzoni:1994ap}
\bibinfo{author}{\bibfnamefont{P.}~\bibnamefont{Marenzoni}},
  \bibinfo{author}{\bibfnamefont{G.}~\bibnamefont{Martinelli}},
  \bibnamefont{and} \bibinfo{author}{\bibfnamefont{N.}~\bibnamefont{Stella}},
  \bibinfo{journal}{Nucl. Phys.} \textbf{\bibinfo{volume}{B455}},
  \bibinfo{pages}{339} (\bibinfo{year}{1995}), \eprint{hep-lat/9410011}.

\bibitem[{\citenamefont{Greensite and Olejnik}(2008)}]{Greensite:2007ij}
\bibinfo{author}{\bibfnamefont{J.}~\bibnamefont{Greensite}} \bibnamefont{and}
  \bibinfo{author}{\bibfnamefont{S.}~\bibnamefont{Olejnik}},
  \bibinfo{journal}{Phys. Rev.} \textbf{\bibinfo{volume}{D77}},
  \bibinfo{pages}{065003} (\bibinfo{year}{2008}), \eprint{arXiv:0707.2860}.

\end{thebibliography}

\end{document}